\documentclass[twocolumn,times]{aastex63}

\usepackage{subfigure}

\definecolor{malachite}{rgb}{0.04, 0.85, 0.32}

\usepackage{soul,xcolor}
\setstcolor{red}

\begin{document}

\shorttitle{AGN-Driven Outflows in the Seyfert 1 Galaxy NGC 3227 on Parsec to Kiloparsec Scales}



\title{An Analysis of AGN-Driven Outflows in the Seyfert 1 Galaxy NGC 3227}

\correspondingauthor{Julia Falcone}
\email{jfalcone2@gsu.edu}

\author[0000-0001-7238-7062]{Julia Falcone}
\affiliation{Department of Physics and Astronomy, Georgia State University, 25 Park Place, Atlanta, GA 30303, USA}

\author[0000-0002-6465-3639]{D. Michael Crenshaw}
\affiliation{Department of Physics and Astronomy, Georgia State University, 25 Park Place, Atlanta, GA 30303, USA}

\author[0000-0002-3365-8875]{Travis C. Fischer}
\affiliation{AURA for ESA, Space Telescope Science Institute, 3700 San Martin Drive, Baltimore, MD 21218, USA}

\author[0000-0001-8658-2723]{Beena Meena}
\affiliation{Space Telescope Science Institute, 3700 San Martin Drive, Baltimore, MD 21218, USA}

\author[0000-0002-4917-7873]{Mitchell Revalski}
\affiliation{Space Telescope Science Institute, 3700 San Martin Drive, Baltimore, MD 21218, USA}

\author[0009-0005-3001-9989]{Maura Kathleen Shea}
\affiliation{Department of Physics and Astronomy, Georgia State University, 25 Park Place, Atlanta, GA 30303, USA}

\author[0000-0003-0483-3723]{Rogemar A. Riffel}
\affiliation{Departamento de Física, CCNE, Universidade Federal de Santa Maria, 97105-900 Santa Maria, RS, Brazil}

\author[0000-0003-3401-3590]{Zo Chapman}
\affiliation{College of Computer Science, Georgia Institute of Technology, 266 Ferst Drive, Atlanta, GA 30332, USA}

\author[0000-0002-7130-7099]{Nicolas Ferree}
\affiliation{Department of Physics, Stanford University, 382 Via Pueblo Mall, Stanford, CA 94305, USA}

\author[0000-0002-2713-8857]{Jacob Tutterow}
\affiliation{Department of Physics and Astronomy, Georgia State University, 25 Park Place, Atlanta, GA 30303, USA}

\author[0009-0005-2145-4647]{Madeline Davis}
\affiliation{Department of Physics and Astronomy, College of Charleston, 66 George Street,
Charleston, SC 29424, USA}



    
\begin{abstract}

We have characterized the ionized, neutral, and warm molecular gas kinematics in the Seyfert 1 galaxy NGC 3227 using observations from the Hubble Space Telescope Space Telescope Imaging Spectrograph, Apache Point Observatory's Kitt Peak Ohio State Multi-Object Spectrograph, Gemini-North's Near-Infrared Integral Field Spectrometer, and the Atacama Large Millimeter Array. We fit multiple Gaussians to several spatially-resolved emission lines observed with long-slit and integral-field spectroscopy and isolate the kinematics based on apparent rotational and outflowing motions. We use the kinematics to determine an orientation for the bicone along which the outflows travel, and find that the biconical structure has an inclination of $40 ^{+5}_{-4}$\arcdeg\ from our line of sight, and a half-opening angle with an inner and outer boundary of $47 ^{+6}_{-2}$\arcdeg and $68 ^{+1}_{-1}$\arcdeg, respectively. We observe ionized outflows traveling 500 km s$^{-1}$ at distances up to 7$''$ (800 pc) from the SMBH, and disturbed ionized gas up to a distance of 15$''$ (1.7 kpc). Our analysis reveals that the ionized outflows are launched from within 20 pc of the SMBH, at the same location as a bridge of cold gas across the nucleus detected in ALMA CO(2-1) observations. We measure a turnover radius where the gas starts decelerating at a distance of $26 \pm 6$ pc from the AGN. Compared to a turnover radius in the range of $31- 63$ pc from a radiative driving model, we confirm that radiative driving is the dominant acceleration mechanism for the narrow line region (NLR) outflows in NGC 3227.

\end{abstract}



\section{Introduction}

At the core of every massive galaxy likely lies a supermassive black hole (SMBH), which has a mass of $10^5 - 10^9$ M$_\odot$. While we observe most SMBHs to be rather quiescent in the nearby Universe, a minority (5$-$10\%) of galaxies possess AGN, wherein the central SMBHs are actively accreting matter and gaining mass from a surrounding accretion disk. Beyond the accretion disk lies a torus of dusty gas that can obscure our vision of the central AGN, depending on its orientation relative to our line of sight (LOS), i.e. unobscured in Type 1, obscured in Type 2 AGN. 
As matter falls from the accretion disk into the SMBH, dynamical interactions between the infalling matter and a loss of potential energy result in the release of massive amounts of radiation. This radiation spans the entire electromagnetic spectrum, often rivaling its host galaxy in brightness.

We utilize our understanding of the AGN kinematics to examine AGN feedback, by which energy and momentum ejected from the nuclear region are deposited into the interstellar medium and incorporated into outflowing material. These outflows are accelerated by AGN radiation pressure and/or magnetic fields \citep{fabian12}, and can result in the expulsion and heating of large quantities of gas, thereby limiting and potentially truncating star formation in the galaxy \citep{fischer17, fischer18, revalski21, venturi21}. Thus we can observe the relationship between the transportation of outflowing material, known as AGN winds \citep{fabian12}, and its effects on surrounding matter, known as AGN feedback. The impact that an AGN feedback cycle can have on the AGN and its host galaxy can be substantial \citep{booth09, angles17} and is associated with the regulation of SMBH accretion, galaxy evolution, and potentially both star formation cessations and bursts \citep{piotrowska21}. Kinematic and geometric models of observations show that the ionized AGN-driven winds flow outward along a biconical geometry, with the central vertex of the bicone intersecting the AGN \citep{antonucci85, pedlar93, nelson00, pogge88, travisthesis}. Spectroscopic subarcsecond observations of the AGN's narrow-line region (NLR) showing blueshifted and redshifted emission allow us to determine the orientation of the cones pointing toward and away from us \citep{travisthesis}. These outflows are often observed in the UV \citep{crenshaw05, crenshaw09}, X-ray \citep{falcao21, breedt10}, optical \citep{colbert96, travisthesis}, and IR wavelengths \citep{fischer17, storchi10, riffel23, vivianu22}.

Many of these observations focus on Seyfert galaxies, which refer to a relatively moderate bolometric luminosity ($L_{bol} \approx 10^{43}-10^{45}$ erg s$^{-1}$), nearby ($z \leq 0.1$) subset of AGN. The Seyfert 1 and Seyfert 2 subclassifications are essential for further characterizing the AGN based on the presence of broad-line and narrow- line regions \citep{khachikian74}. The broad-line region (BLR) consists of dense clouds of ionized gas ($\mathrm{n_H = 10^8-10^{12}}$ cm$^{-3}$) a few light days ($\approx$ 0.005 pc) from the SMBH. Beyond the BLR lies the narrow-line region (NLR), which can extend to a few kpc from the AGN and is composed of significantly more diffuse gas ($\mathrm{n_H \leq 10^6}$ cm$^{-3}$; \citealp{revalski22}). These two regions can be spectroscopically differentiated by observing the emitted lines: the BLR is characterized by the presence of broadened permitted lines because the enormous gravitational influence of the SMBH accelerates the gas to extraordinary velocities. As a result, the emission line becomes significantly Doppler broadened (full width at half maximum [FWHM] = 800 $-$ 8000 km s$^{-1}$). By contrast, the NLR contains both permitted and forbidden line transitions and moves at slower speeds due to its farther distance from the SMBH (FWHM $\leq$ 300 km s$^{-1}$ for rotating gas, but can be as high as 2000 km s$^{-1}$ for outflowing gas; \citealp{fischer18}).
 
Seyfert 1 galaxies have torii oriented in a face-on direction relative to us so that we see ``down the barrel'' to observe both the NLR and the BLR.  By contrast, the orientation of the torii obscures the central engine in Seyfert 2 galaxies in the optical regime, blocking our view of the BLR and resulting in the observation of only the NLR \citep{antonucci93}. 

\begin{figure*}[t]
  \centering
  \subfigure[]{\includegraphics[width=0.4\textwidth]{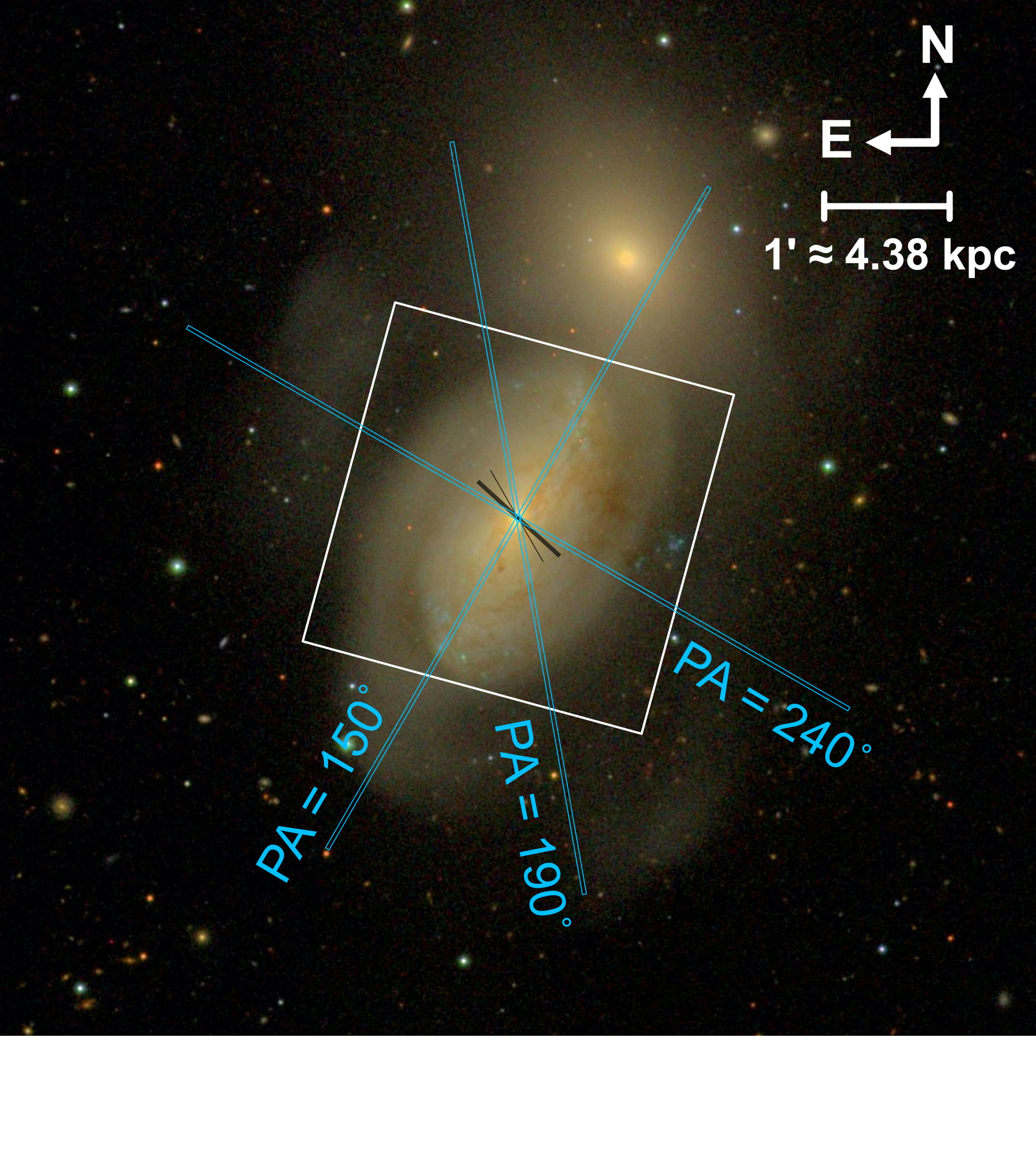}\label{fig:ngc 3227}}
  \hfill
  \subfigure[]{\includegraphics[width=0.57\textwidth]{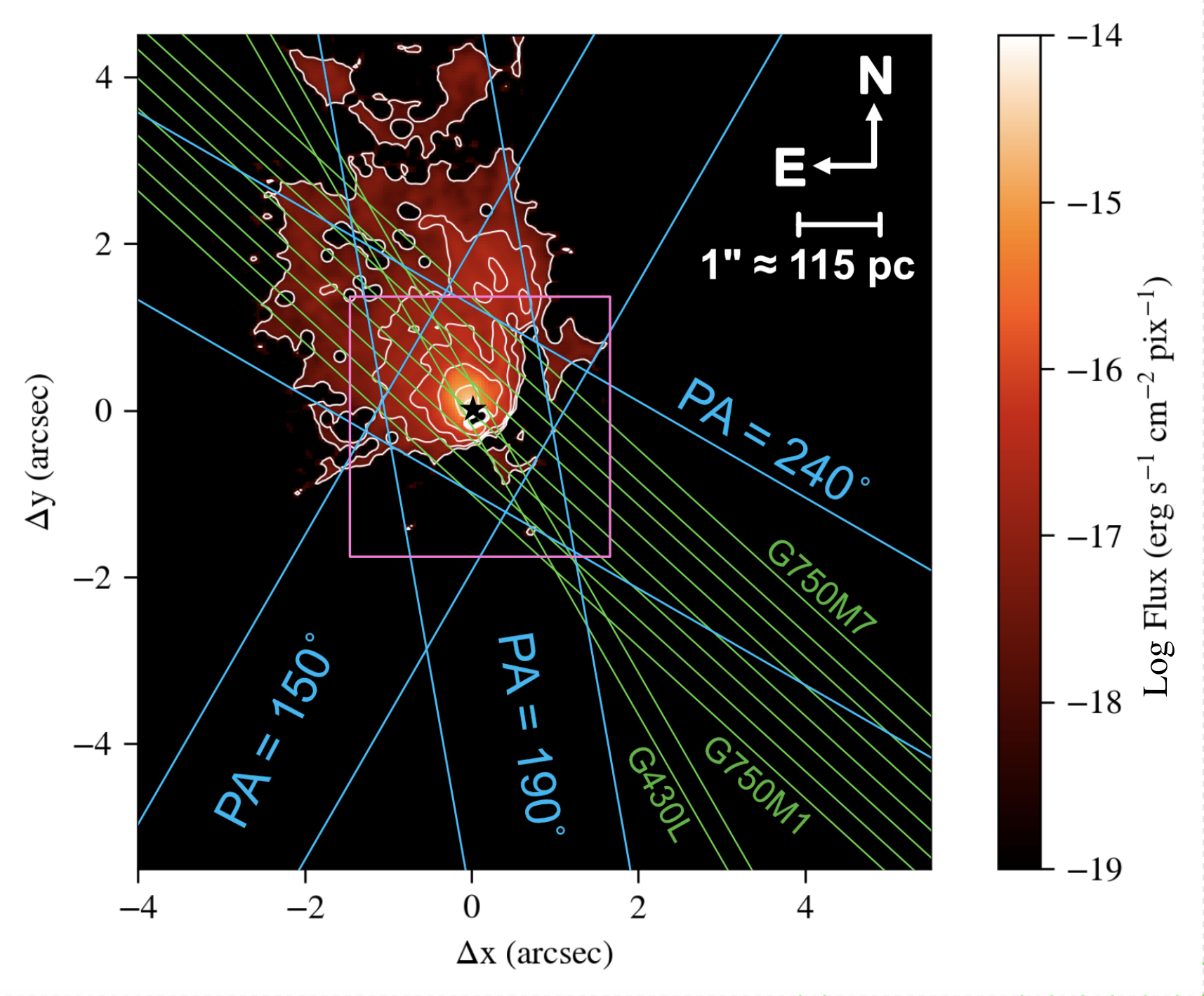}\label{fig:nucleus}}
  \caption{Left: $7.8 \arcmin \times 7.8 \arcmin$ optical image of NGC 3227 (center) and NGC 3226 (top right) taken by the Sloan Digital Sky Survey (SDSS) with \textit{ugriz} filters. The white square shows the position of the WFC3 images. The blue rectangles show the positions of the APO KOSMOS slits, and the outline of the HST STIS slits are shown in black. Right: a continuum-subtracted $9.5\arcsec \times 9.5\arcsec$ \textit{HST} WFC3 F502N image of the nucleus, which shows small-scale structure in the form of [O III] $\lambda$5007 emission. The pink square shows the position of the NIFS field of view. The blue rectangles again show the KOSMOS slits, while the green rectangles give greater detail for the positions of the STIS slits. The seven green parallel STIS slits use the G750M filter, while the slightly tilted green slit uses the G430L filter. The outermost G750M slits are labeled, and the inner slit numbers increase accordingly. The black star shows the location of the bright Seyfert 1 continuum source (i.e., the AGN). }
  \label{fig:fig1}
\end{figure*}
\subsection{NGC 3227}
NGC 3227 ($z$ = 0.003859, pictured in Figure \ref{fig:ngc 3227}) is a Seyfert 1, SAB(s)a type galaxy \citep{deVaucouleurs91}. It is located in the constellation Leo at a distance of $23.7 \pm 2.6$ Mpc \citep{tonry01, blakeslee01} so that its transverse scale corresponds to $\sim$115 pc arcsec$^{-1}$ on the plane of the sky. NGC 3227 has a complex dynamical history through its interactions with its companion NGC 3226. \cite{rubin68} documented the galaxy pair in depth, using long-slit spectroscopy to identify nuclear gas outflows and an arm extending from NGC 3227 to NGC 3226 at a mean velocity of 550 km s$^{-1}$ with respect to the NGC 3227/3226 system. \cite{mundell95} followed up with neutral H I imaging to detect two tidal tails on NGC 3227, one of which extends $\sim$70 kpc northward towards and beyond NGC 3226, and the other extending $\sim$30 kpc southward. They also discovered J1023+1952, an H~I-rich dwarf galaxy about $40''$ wide located in front of the western side of NGC 3227's disk. 


\begin{table*}[tt]
\centering
\footnotesize
\begin{tabular} 
{|c c c c c c c c c c c|} 

 \hline

 Instru-	& Program& Observation	& Date 	& Grating/	& Exposure & Wavelength 	& Spectral	&Spatial  &	Position  &Spatial \\
 
ment& ID & or Program& (UT)  & Filter/	& Time 	&  Range 	&  Dispersion 	& Scale  &Angle & Offset  \\

&  & ID & & Grism & (s)	&  (\AA)&   (\AA/pix)	&  ($''$/pix) & (deg) &($''$ NW) \\

 \hline
STIS & 7403 & O57204010 & 1999 Jan 31 & G750M & 2105 & $6248-6912$ & 0.56 & 0.051 & -137.62 & -0.75 \\
STIS & 7403 & O57204020 & 1999 Jan 31 & G750M & 1600 & $6248-6912$ &  0.56 & 0.051 & -137.62 & -0.5 \\
STIS & 7403 & O57204030 & 1999 Jan 31 & G750M & 1884 & $6248-6912$ &  0.56 & 0.051 & -137.62 & -0.25 \\
STIS & 7403 & O57204040 & 1999 Jan 31 & G750M & 1890 & $6248-6912$ &  0.56 & 0.051 & -137.62 & 0 \\
STIS & 7403 & O57204050 & 1999 Jan 31 & G750M & 1600 & $6248-6912$ &  0.56 & 0.051 & -137.62 & 0.25 \\
STIS & 7403 & O57204060 & 1999 Jan 31 & G750M & 1884 & $6248-6912$ &  0.56 & 0.051 & -137.62 & 0.5 \\
STIS & 7403 & O57204070 & 1999 Jan 31 & G750M & 1887 & $6248-6912$ &  0.56 & 0.051 & -137.62 & 0.75 \\
STIS & 8497 & O5KP01040 & 2000 Feb 08 & G750L & 120 & $2900-5700$ & 4.92 & 0.051 & -150.34 & 0 \\
STIS & 8479 & O5KP01020 & 2000 Feb 08 & G430L & 120 & $2900-5700$ & 2.73 & 0.051 & -150.34 & 0 \\
WFC3 & 16246 & iebn12020 & 2020 Oct 28 & F502N & 1332 & $4972-5049$ & N/A & 0.039 & -20 & N/A \\
WFC3 & 16246 & iebn12010 & 2020 Oct 28 & F547M & 720 & $5075-5866$ & N/A & 0.039 & -20 & N/A \\
KOSMOS & GS01 & N/A & 2021 Dec 04 & Blue & 2700 & $4150-7050$ & 0.71 & 0.257 & 150 & 0 \\
KOSMOS & GS01 & N/A & 2021 Dec 04 & Blue & 2700 & $4150-7050$ & 0.71 & 0.257 & 240 & 0 \\
KOSMOS & GS01 & N/A & 2022 Jan 29 & Blue & 2700 & $4150-7050$ & 0.71 & 0.257 & 190 & 0 \\
NIFS & N/A &  GN-2018B-Q-109 & 2018 Dec 13 & Z & 1800 & $9400-11500$ & 1.03 & 0.043 & 0 &0\\
NIFS & N/A &  GN-2016A-Q-6 & 2016 Feb 28 &J  & 2400 & $11500-13300$ & 0.88 & 0.043 & 0 &0\\
NIFS & N/A &  GN-2016A-Q-6 & 2016 Feb 28 & K& 2400 & $19900-24000$ & 2.09 & 0.043 & 0 & 0\\

 \hline

\end{tabular}
\caption{HST, KOSMOS, and NIFS observations of NGC 3227. The columns list (1) instrument, (2) Program ID, (3) Observation ID, (4) observation date, (5) grating (for spectra) or filter (for imaging), (6) exposure time for each observation, (7) wavelength range (of the spectra) or bandpass (of the filter), (8) spectral dispersion of the grating, (9) spatial scale of the spectra, (10) position angle of the slits or angle of the images, using east-of-north orientation, and (11) the spatial offsets of the slits or images from the nucleus. The values for columns (8)-(9) for the HST data were obtained in their corresponding instrument handbooks \citep{dressel12, riley17}.}
\label{table:obs}
\end{table*}

NGC 3227 is a dynamic and fascinating Seyfert 1 galaxy from which, even after decades of study, we can learn valuable information about AGN feedback and outflow processes. There have been previous efforts by \cite{travisthesis} to constrain the orientation of the biconical geometry of the outflows by measuring narrow and broad emission lines in the NLR of NGC 3227. Whereas \cite{travisthesis} relied primarily on optical data from the Hubble Space Telescope (HST) Space Telescope Imaging Spectrograph (STIS) instrument, we add multiphase and multiscale observations comprising optical, near-IR, and radio data that exhibit kinematic details on both nuclear and galaxy-wide scales. Additionally, we have made significant improvements to the methodology of emission line fitting and the subsequent determination of the best geometric model for the data. Those improvements and the resulting model are described in this paper. Finally, the NGC 3227/3226 system has a significant history of interaction. We are interested in examining whether or not a companion galaxy affects a host galaxy's feeding and feedback processes, and plan to address this phenomenon in future papers.

\section{Observations}\label{sec2}

\subsection{Hubble Space Telescope (HST)}
\subsubsection{Space Telescope Imaging Spectrograph (STIS)}

We used HST STIS archival observations to probe the kinematics of NGC~3227 on small scales. The medium dispersion grating G750M has a spectral resolution of 1.8 \r{A} (FWHM $\sim$ 83 km s$^{-1}$ at 6500 \r{A}) and a spatial resolution of $\sim$0\farcs10 (2 pixels for Nyquist sampling). As shown in Figure \ref{fig:nucleus}, seven parallel 52$''$ $\times$ 0\farcs2 G750M slit observations, each separated by a 0\farcs05 gap and with position angle (PA) = -137.62\arcdeg, were taken with Hubble program ID 7403 (PI: A. Filippenko). The lower dispersion grating G430L has a spectral resolution of 9.0 \r{A} FWHM ($\sim$540 km s$^{-1}$) and the same spatial resolution as the G750M. A single slit observation at PA = -150.34\arcdeg was taken with Hubble program ID 8479 (PI: S.B. Kraemer), which included a G750L spectrum at the same slit position as the G430L slit, but is not used in this paper. The middle G750M slit (observation ID: O57204040) and the G430L slit are centered over the AGN. A summary of the HST data is shown in Table \ref{table:obs}. We retrieved the calibrated data from the Mikulski Archive at the Space Telescope Science Institute (MAST), and then used Interactive Data Language (IDL) to extract the two-dimensional spectral images and convert to flux per cross-dispersion (0\farcs05 $\times$ 0\farcs2) element.

\subsubsection{Wide Field Camera 3 (WFC3)}
We connected the locations of distinct NLR knots of emission to their measured kinematics  using [O~III] images from WFC3 observations, which can be used to trace the ionized gas outside the spatial range covered by the STIS slits. We used a F547M image as the continuum to subtract from the line emission of the F502N image. Both images were obtained from HST program ID 16246 (PI: M. Revalski).
This continuum subtraction first involved a conversion of data from electrons per second to erg cm$^{-2}$ s$^{-1}$ for both images, which is done by multiplying the data counts for each image by its filter's inverse sensitivity and its filter bandwidth. The resulting image exhibiting the ionized gas is shown in Figure~\ref{fig:nucleus}.

\subsection{Kitt Peak Ohio State Multi-Object Spectrograph (KOSMOS)}

Using the Kitt Peak Ohio State Multi-Object Spectrograph (KOSMOS) instrument at APO, we obtained additional long-slit spectroscopy of NGC 3227 to map the kinematics of the ionized gas on large (galactic) scales. The position angles of these measurements were chosen to maximize coverage of the NLR as well as the major and minor axes of the host galaxy. KOSMOS allows for three possible slit locations that determine the desired wavelength range for a single grism. For all KOSMOS measurements, we used the blue grism located at a ``high" slit location to provide data in the wavelength range of 4150 $-$ 7050 \r{A}. This slit position allowed for the measurement of both [O~III] and H$\alpha$, which are crucial in our analysis. Information on the observations can be found in Table~\ref{table:obs}. 

We reduced the data to obtain calibrated spectral images using standard IRAF routines including bias subtraction and flat-field correction \citep{tody86, tody93}. Standard stars selected from the \cite{oke90} catalog were observed for flux calibration, and arc lamp images taken at the telescope location of each science exposure were used for wavelength calibration. Additional calibrations were performed in IDL to correct the tilt of the spectrum in the cross-dispersion direction \citep{gnilka20} to produce calibrated 2D spectral images.  

\subsection{Gemini North Near-Infrared Field Spectrograph (NIFS)}

We utilized archival observations of NGC 3227 in the J- and K-bands from the Near-Infrared Field Spectrograph (NIFS) at Gemini North. We also took new Z-band observations with NIFS as shown in Table~\ref{table:obs}. These integral field unit (IFU) observations obtained with adaptive optics have an angular resolution of $\sim$0\farcs1 over a 3$''$ $\times$ 3$''$ field of view, which is important for characterizing the outflows near the nucleus. Data reduction included image trimming, flat fielding, sky subtraction, and wavelength calibration. We followed the pipeline outlined in \cite{merrell20}, which improves the standard NIFS data reduction pipeline to account for necessary corrections such as flux calibration and variance propagation to the final data cube. The pipeline described in \cite{merrell20} was only written to reduce the K-band, so we modified the code to include compatibility with the Z- and J-bands. Reduced versions of the same J- and K-band observations were obtained from \cite{riffel17} to verify consistency.

Each data cube covers 3$''$ $
\times$ 3$''$, which equates to 345 pc $
\times$ 345 pc assuming a distance of 23.7 Mpc \citep{tonry01}. The emission lines that are observed in these spectra represent distinct gas phases. Specifically, the H$_2$ emission from the K-band maps the warm molecular gas; Pa$\beta$ and He I are recombination lines in the J- and Z-band, respectively, that trace a broad range of ionized gas; and [S III] emission from the Z-band traces moderately-ionized gas similar to [O~III] in the optical.

\section{Methods}
\label{sec:methods}

\begin{figure*}[!]
  \centering
  \includegraphics[width=0.49\linewidth]{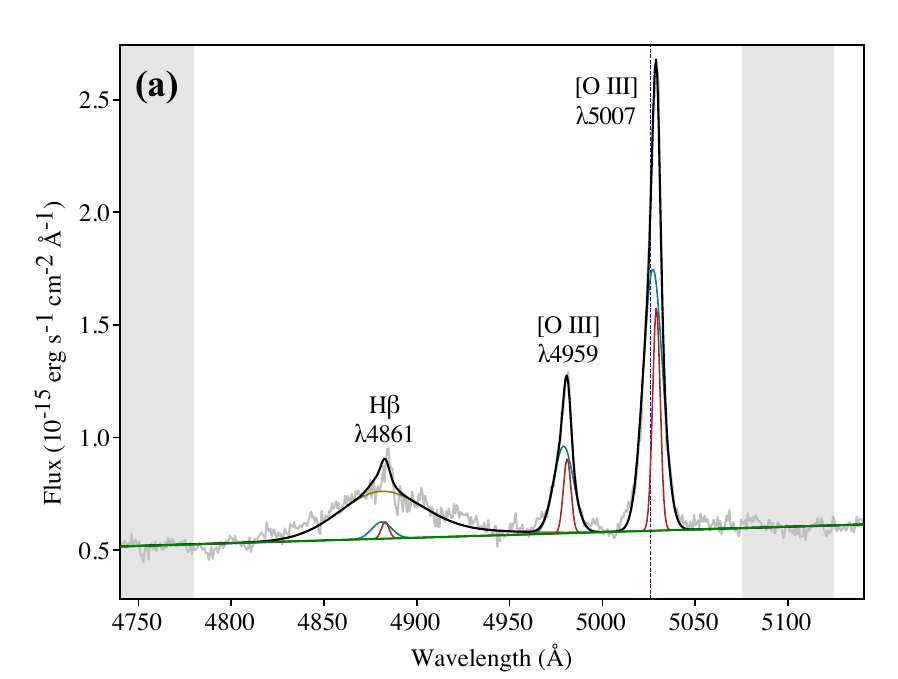}
\includegraphics[width=0.49\linewidth]{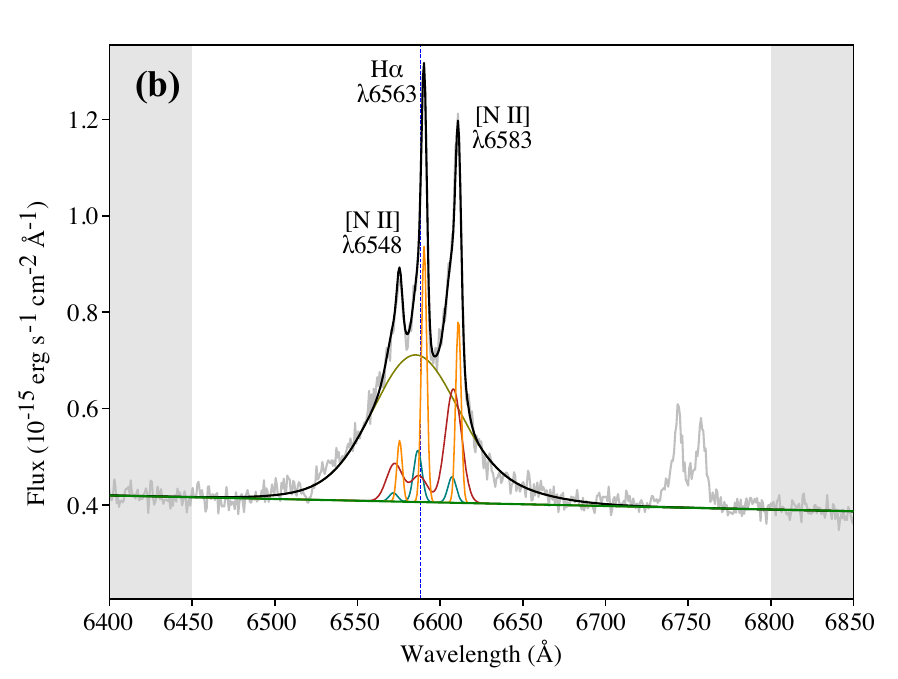}
  \caption{Examples of BEAT fitting (a) H$\beta$ $\lambda$4861 and [O III] $\lambda\lambda$4959/5007 emission lines and (b) H$\alpha$ $\lambda$6563 and [N II] $\lambda\lambda$6548/6583 emission lines using KOSMOS data. The blue dashed line shows the systemic redshift of the H$\alpha$ and [O III] $\lambda$5007  lines, respectively. We plot the spectra, systemic redshift, and fits in the observed wavelength frame. The double-peaked feature at $\sim$6750 \r{A} is the [S II] $\lambda$6716/6731 doublet, which will be incorporated into future fits to obtain gas densities. In both figures, the black curve is the composite fit, while the gray curve is the observed spectrum. The olive curve is the broad component while the red, orange, and teal curves are the narrow kinematic components. The green line marks the continuum level, which is a linear fit to the averaged fluxes contained within the gray areas to the sides of the emission lines.}
  \label{fig:beat_curves}
\end{figure*}

To understand the NLR kinematics of NGC 3227, we used the long-slit spectra to isolate two groups of emission lines. The first group consists of H$\beta$ $\lambda$4861 and [O III] $\lambda \lambda$4959/5007, while the second consists of H$\alpha$ $\lambda$6563 and [N II] $\lambda \lambda$6548/6583. Figure~\ref{fig:beat_curves} shows these emission lines in a spectrum taken from the nuclear region. Whereas KOSMOS can record emission lines from both groups in the same observation, we have HST data utilizing the G430L and G750M slits to cover the first and second group, respectively. 

\subsection{Bayesian Evidence Analysis Tool (BEAT)}
The 2D spectra allow us to characterize changes in emission lines as we look away from the nucleus. We quantified these changes at each position along the slit by fitting multi-component Gaussian profiles using the Bayesian Evidence Analysis Tool (BEAT; \citealp{fischer17}) routine, which can separate kinematic components. Physically, these components are tied to filaments and knots like those seen in Figure \ref{fig:nucleus}. While the STIS slits can resolve individual filaments, the wide area covered by the KOSMOS slit means that the resulting spectrum at each point along the slit is a superposition of multiple knots' motions. As a consequence, we expect and often find that areas closer to the AGN, which tend to be more filamentary, will be fit with more components (representing the individual knots) than areas that are farther from the nucleus. 

\begin{figure*}[t]
\centering  
 \includegraphics[width=0.32\linewidth]{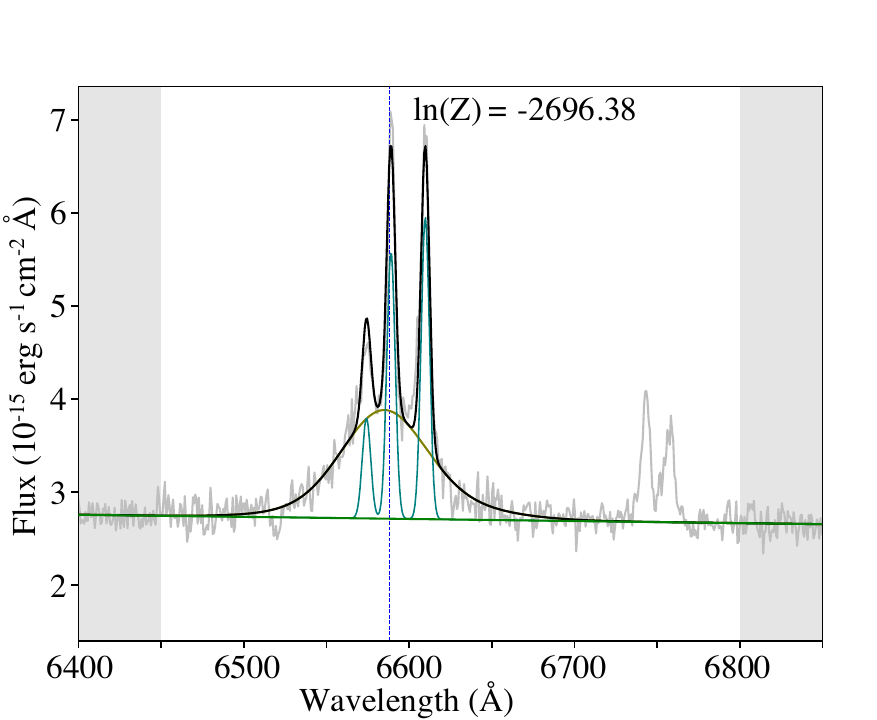}
\includegraphics[width=0.32\linewidth]{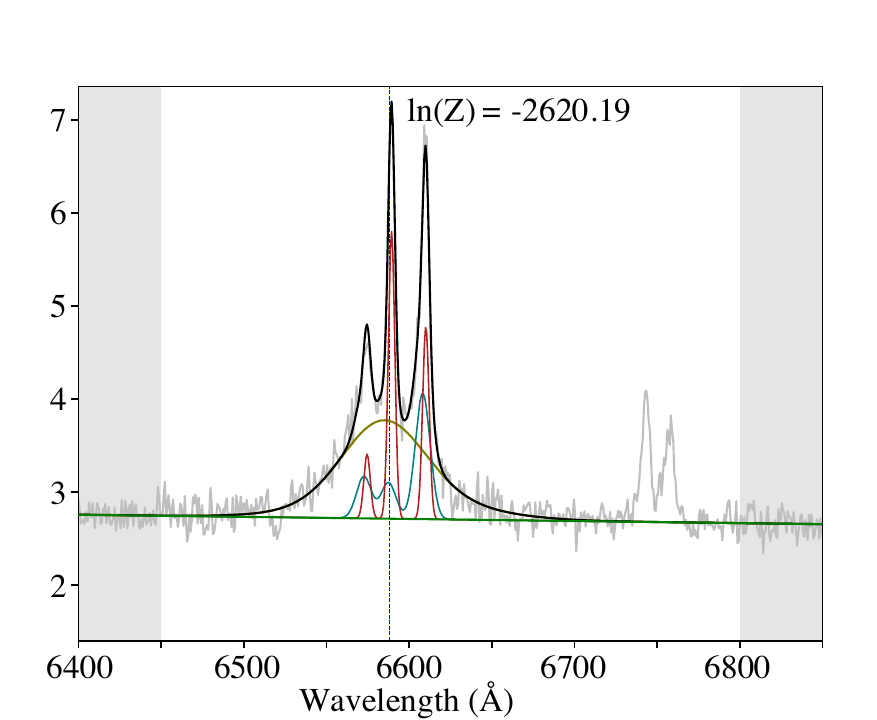}
 \includegraphics[width=0.32\linewidth]{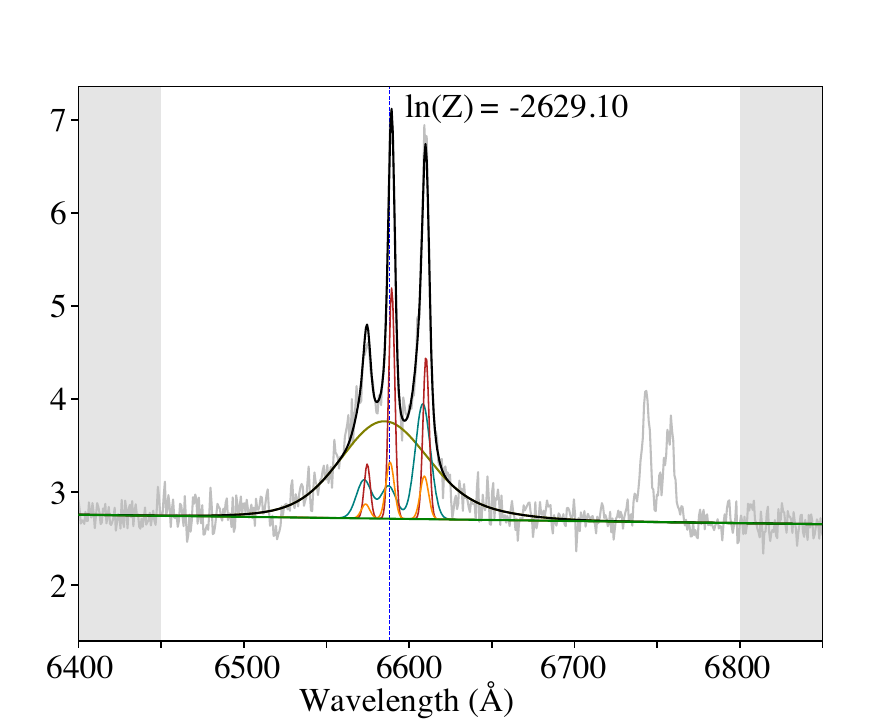}
\caption{The result of BEAT fitting one (left), two (middle), and three (right) narrow-line components to a spectrum. The broad line component (olive) is present in all three fits. BEAT determined that for this spectrum, the two component fit was optimal.}
\label{fig: beat comps}
\end{figure*}

We applied BEAT to spectra from both long-slit spectroscopy (STIS and KOSMOS data) and IFU data cubes (NIFS data), fitting each spectrum in the slit and data cube, respectively. BEAT utilizes the MultiNest sampling algorithm \citep{feroz08, feroz09, feroz19, buchner14} to compute Bayesian statistics and determine the fewest number of Gaussian components that significantly improve the fit of the model to the data. For a set of predetermined narrow emission-line species and a maximum number of allowed Gaussian kinematic components, BEAT calculates the likelihood of a model. 
Although emission line profiles are not always purely Gaussian \citep{heckman81, veilleux91}, treating them as such for simplicity and efficiency has been successfully employed for this type of study and has allowed for accurate determination of velocity centroids and widths, as well as fluxes of the emission-line components \citep{revalski18, revalski21, fischer18, gnilka20}. 
Gaussian model parameters were fixed within a given range, including the intrinsic velocity widths, acceptable wavelengths for line centroids, and the relative intensities of certain lines such as [O~III] $\lambda$5007/4959 and [N~II] $\lambda$6583/6548 as 3.01 and 2.95, respectively \citep{osterbrock06}. 
The lower ends of the line widths were restricted by the instrument’s spectral resolution. Although other kinematic studies of the NLR \citep{crenshaw00, travisthesis, fischer18, meena23} show outflows on scale of 1000-2000 km s$^{-1}$, we set the upper limit of the NLR to 1400 km s$^{-1}$ (FWHM), which sufficiently divides the NLR from the BLR in our fits. Consequently, when we fit the BLR of these spectra, we utilized a lower limit of 1400 km s$^{-1}$ (FWHM).  The lower limit for the flux levels of each component is 3$\sigma$ of the continuum noise, and the upper limit is given by the peak flux of the corresponding emission line data. We do not fit data that fail the criterion of S/N $\ge$ 3.

BEAT also fits the continuum and allows for the specification of the underlying broad-line emission, the process of which for NGC 3227 is described in Section \ref{sec: broad lines}. It iterates through the parameter spaces, fitting Gaussians and calculating their likelihoods, returning the likeliest model as the best fit. For a given model, the likelihoods from all fits are summed to give a Bayesian evidence value \textit{\textbf{Z}}. The evidence value of each component describes the overall likelihood of the number of components corresponding to the observed spectrum. The logarithm of \textit{\textbf{Z}} is taken, and shown in each panel of Figure \ref{fig: beat comps}. The log(\textit{\textbf{Z}}) value closest to 0 is representative of the most likely number of components for the spectrum unless its improvement is insignificant relative to the previous model fit. In other words, going from one number of components to the next, 
\begin{equation}\label{my_first_eqn}
\mathrm{ln}\left[\frac{Z_{i+1}}{Z_{i}}\right] \geq 5
\label{eq1}
\end{equation}
as per \cite{feroz11} where $Z_i$ and  $Z_{i+1}$ are the evidence values for two models fit with \textit{i} and \textit{i}+1 components, respectively. If Equation \ref{eq1} is true, then that constitutes “strong evidence" (99.3\% probability) that the more complex model is superior \citep{feroz11}. Figure \ref{fig: beat comps} depicts an example fit where the evidence values reveal that the two-component model is the optimal choice for this spectrum.



Wavelength centroid, line width, and peak flux values of each component from our best fits are used to create kinematic plots that show the radial velocity, FWHM, and integrated flux profiles of the individual kinematic components. These plots, described in greater detail in Section \ref{kinematics}, are integral to interpreting the gas kinematics.


\subsection{Fitting Broad Lines}
\label{sec: broad lines}
A defining characteristic of Seyfert 1 AGN like NGC 3227 is the presence of broad emission lines in the nuclear region. Although the narrow components can vary along the slits according to BEAT’s input parameters, the broad component is unresolved in our observations and is treated as a point source, and thus has fixed kinematics along the slits. As a result, the broad component maintains a constant width and wavelength centroid along the slit, so the only allowed variation is in the flux of the profile as it changes according to the point spread function (PSF). 

For each KOSMOS, STIS, and NIFS observation, we extracted a one-dimensional spectrum from the nuclear region. Specifically, we used the spectrum with the brightest continuum flux as our template for the broad components. We allowed the routine to freely fit the broad emission with multiple Gaussian components, with the stipulation that the line widths must be greater than 1400 km s$^{-1}$ to avoid contamination from narrow line components.
We ascribe no significance to the individual broad components, and some may be due to other emission lines such as Fe~II in the H$\beta$ region.
We performed these fits separately for different epochs of observation because the broad lines in NGC~3227 are known to vary in both profile and flux \citep{bentz23}.
The resulting broad line components with fixed relative intensities are used as templates in all subsequent fits for the H$\beta$ and H$\alpha$ regions. 

\subsection{Error Analysis}
To use BEAT, we must input an uncertainty with each data point on every spectrum we fit. 
For the NIFS data, after completing the pipeline outlined in \cite{merrell20}, we utilized the variance data cube and take the square root of the variance for each point along each spectrum to approximate the noise.  For our KOSMOS and STIS data, we multiplied the standard deviation of flux of the nuclear row by the square root of the flux at each point along the spectrum divided by the average flux of the nuclear row. All three methods were sufficient for producing noise arrays BEAT utilized in its fitting of all the spectra. 

For our STIS and KOSMOS spectra, the errors for the BEAT fits were calculated by manually dividing the data in each slit into distinct kinematic groups based on position and velocity. For each group, we calculated the median absolute deviation and used that value as the velocity uncertainty for each member in that group. In these instances, we calculated our own errors because BEAT's uncertainties were several times higher than expected for our resolved spectra. We did not redo the uncertainties for the NIFS data because they are reasonable values. However, in kinematic plots in this paper, we only display velocity uncertainties for the STIS data because these are the only uncertainties that we use in subsequent analyses in this paper.

\section{Kinematics} \label{kinematics}

\begin{figure*}
\centering  
\subfigure[]{\includegraphics[width=\linewidth]{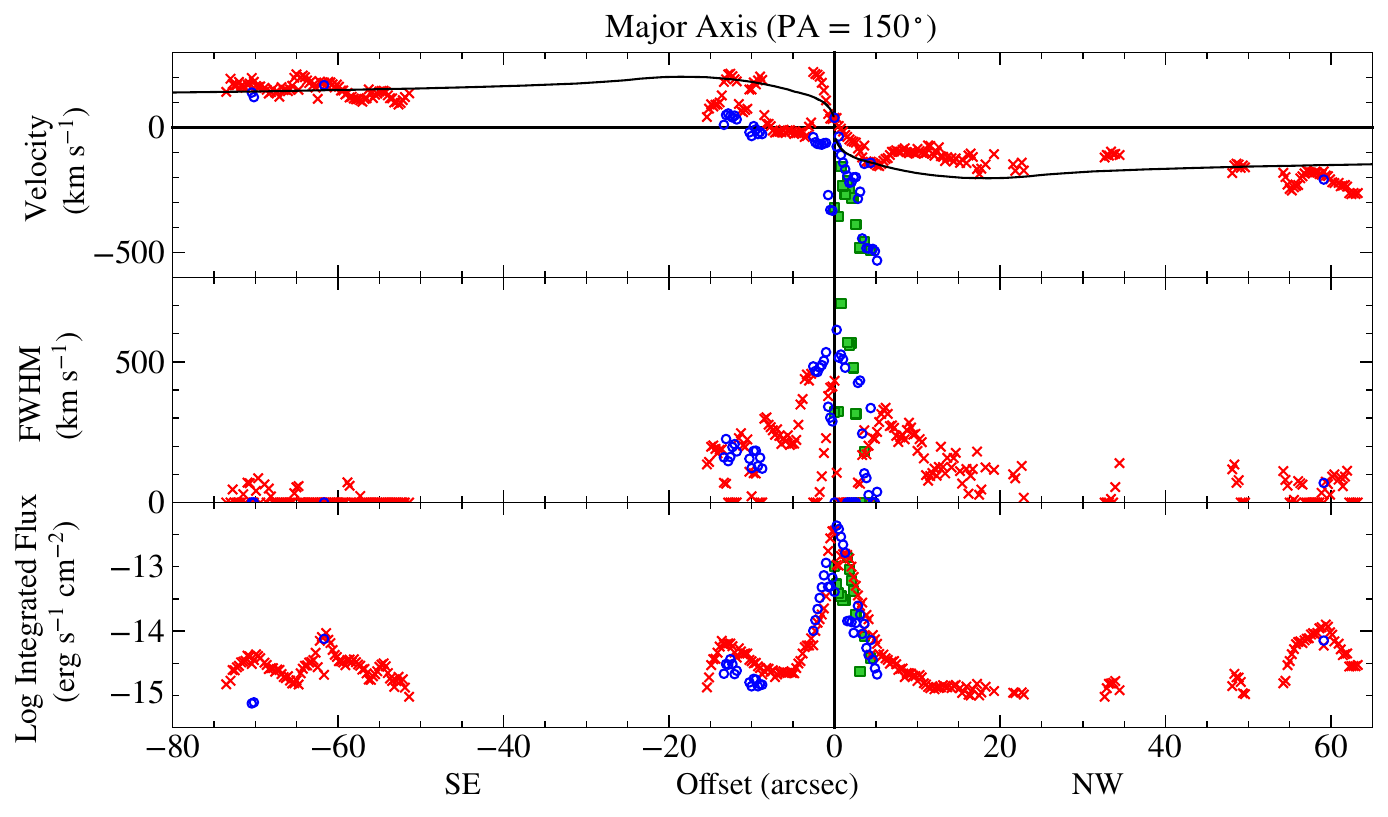} \label{fig: KOSMOS major axis}}
 \vspace{.2cm}

\subfigure[]{\includegraphics[width=0.49\linewidth]{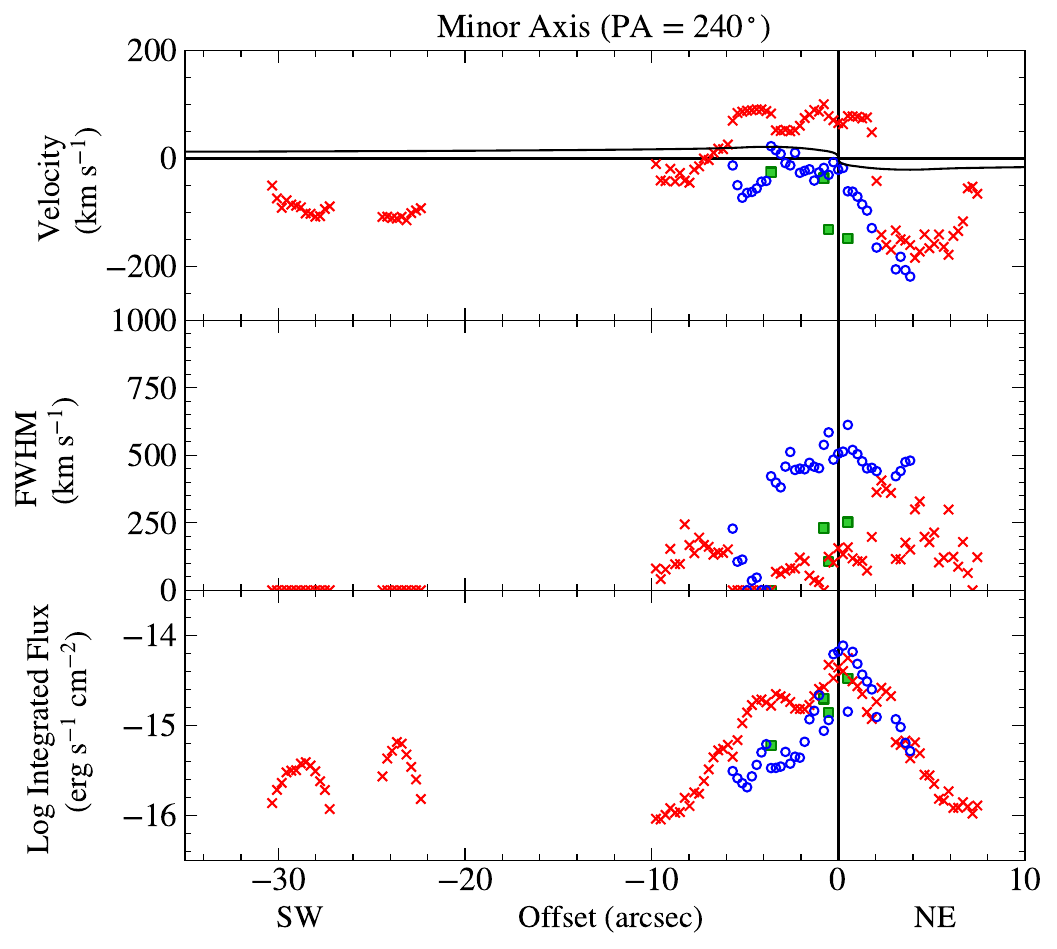}\label{fig: KOSMOS minor axis}}
\subfigure[]{\includegraphics[width=0.49\linewidth]{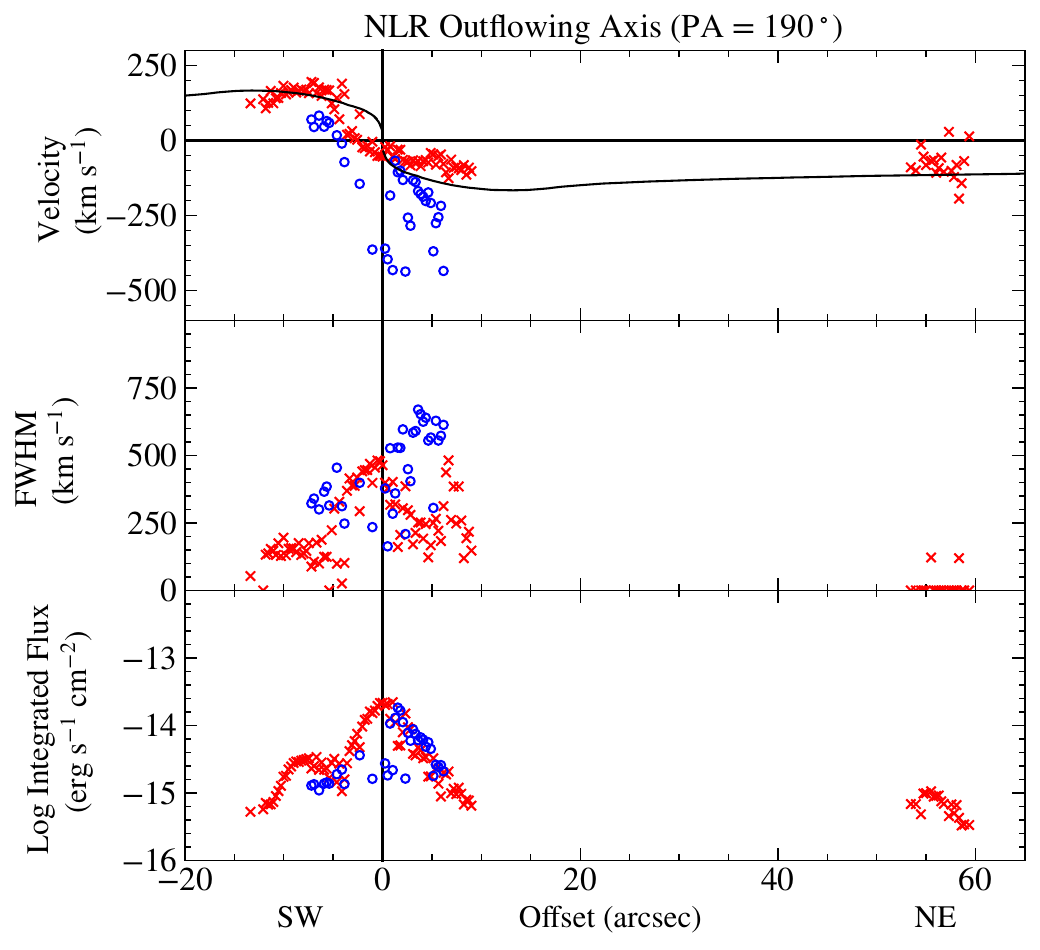}\label{fig: KOSMOS outflow axis}}
\caption{H$\alpha$ kinematic plots for the KOSMOS slits along the major (top) and minor (lower left) axes of the galaxy, and outflow axis of the NLR (lower right). The top panel of each figure shows the radial velocity distribution of H$\alpha$ emission, the middle panel shows the FWHM distribution of each component, and the bottom panel shows the integrated flux distribution. The kinematic components at each location are sorted according to decreasing velocity in red (X), blue (circles), and green (squares).
The black curve running through all velocity plots is the rotation curve of NGC~3227 taken from \cite{schinnerer00}, projected along the slit position angles for the host galaxy parameters given in Section \ref{newmodel}.}
\label{fig: KOSMOS kinematics}
\end{figure*}

\subsection{APO KOSMOS Kinematics}
\label{sec: kosmos kin}
The wide wavelength range of our KOSMOS measurements allow us to measure  H$\beta$ $\lambda$4861, [O~III] $\lambda \lambda$4959/5007, H$\alpha$ $\lambda$6563 and [N~II] $\lambda \lambda$6549/6585 simultaneously. However, in our analysis of KOSMOS data we chose to focus on H$\alpha$ (and by extension, [N~II]) because of its brightness, ability to trace star formation and AGN activity, and because it is less susceptible to extinction effects than the bluer emission lines.

Figure~\ref{fig: KOSMOS kinematics} shows the ionized gas kinematics for H$\alpha$ in our three KOSMOS slits. BEAT returned a maximum of three components for each fit, an upper limit that was established based on observed kinematics in past studies \citep{fischer17, meena21, meena23, revalski21} and to avoid excessive computational time. A consistent pattern in all three slits is that we see the most components near the nucleus (over which the slits are centered), and fewer components in the outer regions. The black curve present in all velocity plots of Figure~\ref{fig: KOSMOS kinematics} originates from the rotation curve produced in \cite{schinnerer00}, which was derived from molecular $^{12}$CO data. The maximum and minimum velocity values, and the extent to which the rotation curve is stretched or squeezed in the spatial direction, vary according to the projection of the rotation curve at each position angle. We can classify the kinematics in the outer ($\ge$ 20$''$, or 2.3 kpc) regions as predominantly rotational motion, as evidenced by the flatness of the velocity profiles which is consistent with the expected flatness of the rotation curve \citep{rubin70}. The fits in these regions primarily comprise a single component, meaning that rotational motion around the galactic disk is the only observable motion. This can be contrasted with the presence of two or three components in the inner  ($\le$ 20$''$) regions of all three slits.

To distinguish outflows from rotation, we have sorted the components in Figure~\ref{fig: KOSMOS kinematics} from high to low velocity. In Figure~\ref{fig: KOSMOS major axis}, we observe velocities and FWHMs along the major axis of the galaxy consistent with an extension of the rotation curve at projected distances of 15$''$ (1.7 kpc) to 75$''$ (8.6 kpc), where the H$\alpha$ emission is primarily from H~II regions. The flux plots in Figures \ref{fig: KOSMOS major axis} and \ref{fig: KOSMOS minor axis} also show hump-like shapes at distances of 55$-$65$''$ and $-$30$''$ to $-$20$''$, respectively. These correspond to the locations of the spiral arms in NGC 3227.

The red points in Figures \ref{fig: KOSMOS major axis} and \ref{fig: KOSMOS outflow axis} generally trace the rotation curves, with a few discrepancies at $-$4$''$ to $-$2$''$ in Figure \ref{fig: KOSMOS major axis} and $-$5$''$ to 5$''$  in Figure \ref{fig: KOSMOS outflow axis}. At these locations, the radial velocities $|v_r|$ are practically 0 km s$^{-1}$, which are substantially lower than the velocities predicted by the rotation curve in those regions of around 200 km s$^{-1}$. Figure~\ref{fig: KOSMOS minor axis} shows minimal agreement between the rotation curve and any of the kinematics, with the observed $|v_r|$ measurements reaching 100$-$200 km s$^{-1}$ despite expected projections of rotational motion of around 10 km s$^{-1}$.
Overall, significant excursions from the rotation curve in all three components indicate outflows in the nuclear region of NGC~3227
that reach projected velocities up to 500 km s$^{-1}$.

\subsection{\textit{HST} STIS Kinematics}
\label{sec: STIS kinematics}

Figure~\ref{fig: STIS kinematics} shows the ionized gas kinematics for the seven STIS G750M slits tracing H$\alpha$ gas, as well as one G430L slit that traces [O~III].
These kinematic plots are in good agreement with those produced by \cite{travisthesis}, where they performed a similar analysis on these data. 
Slight differences arise because they did not use the BEAT fitting routine, but instead used another Gaussian fitting routine wherein the number of components is determined by eye. With the predetermined number of components, the program then chose the model with the minimum chi-squared as the best fit. However, a chi-squared parameter can be lowered by adding additional Gaussian components, which can cause overfitting. Using the Bayesian likelihood to evaluate the best model, as BEAT does, minimizes the chances of overfitting. In instances where we fit the same number of components as them at a given distance along a slit, the kinematics match. Nevertheless, we see variation in the kinematics from them when we fit a different number of components at that location along the slit, or if we fit components at more extended distances along the slit than they could resolve with their routine.

\begin{figure*}[t]
\centering  
\includegraphics[width=0.45\textwidth]{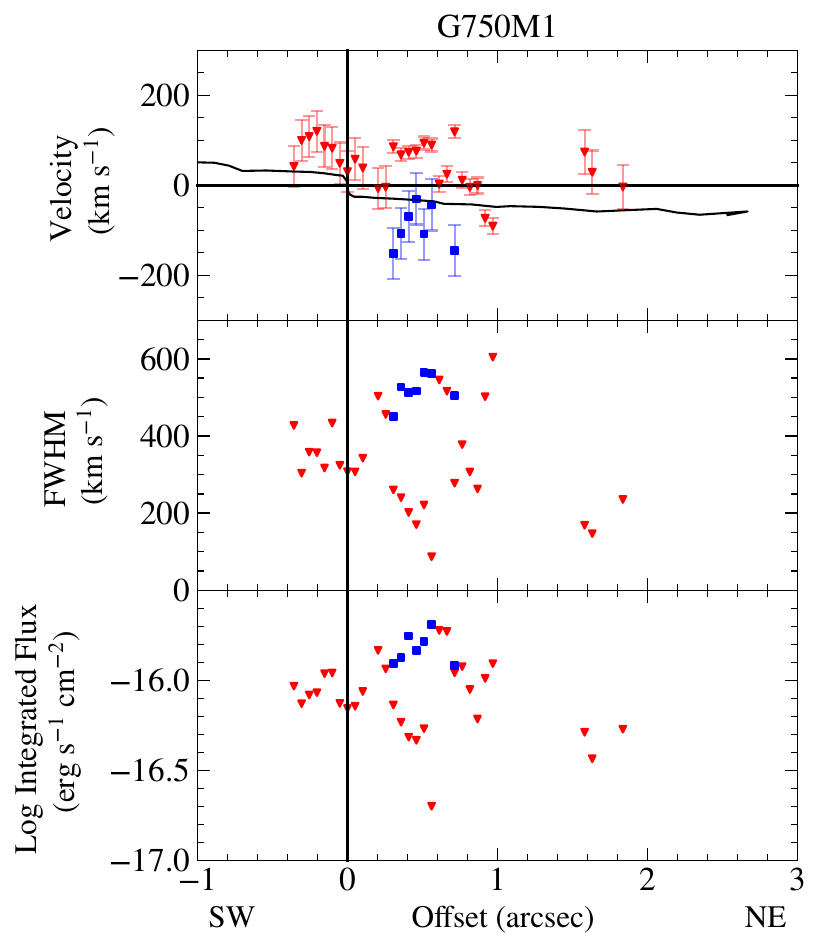}
\includegraphics[width=0.45\textwidth]{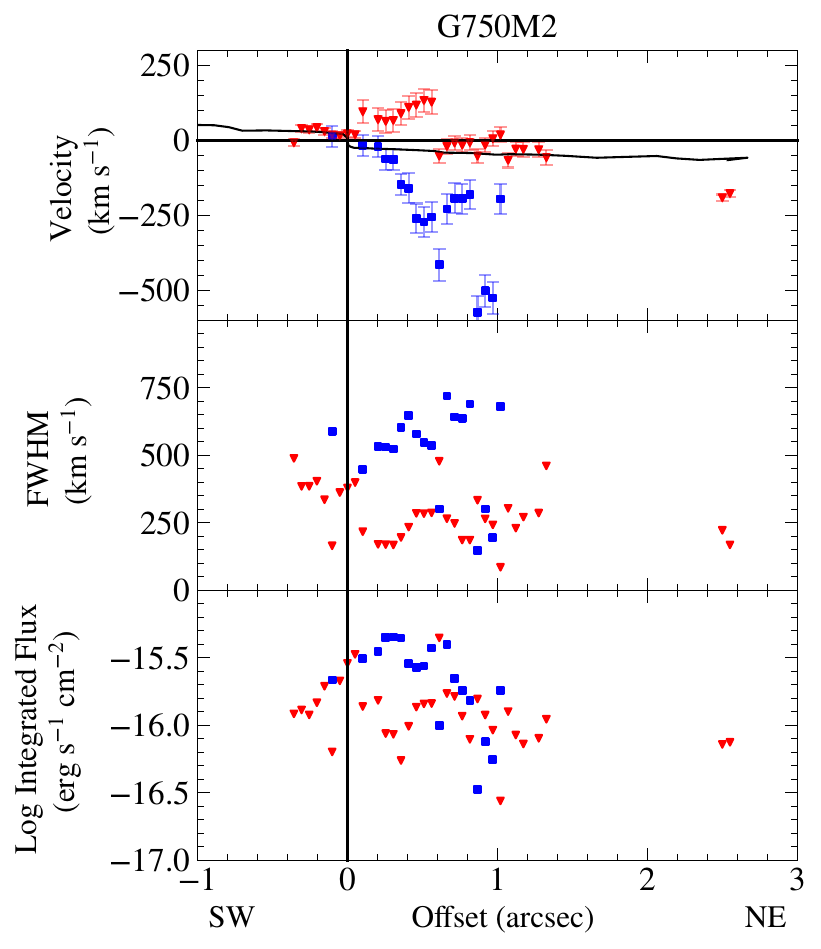}
 \vspace{.25cm}
\includegraphics[width=0.45\textwidth]{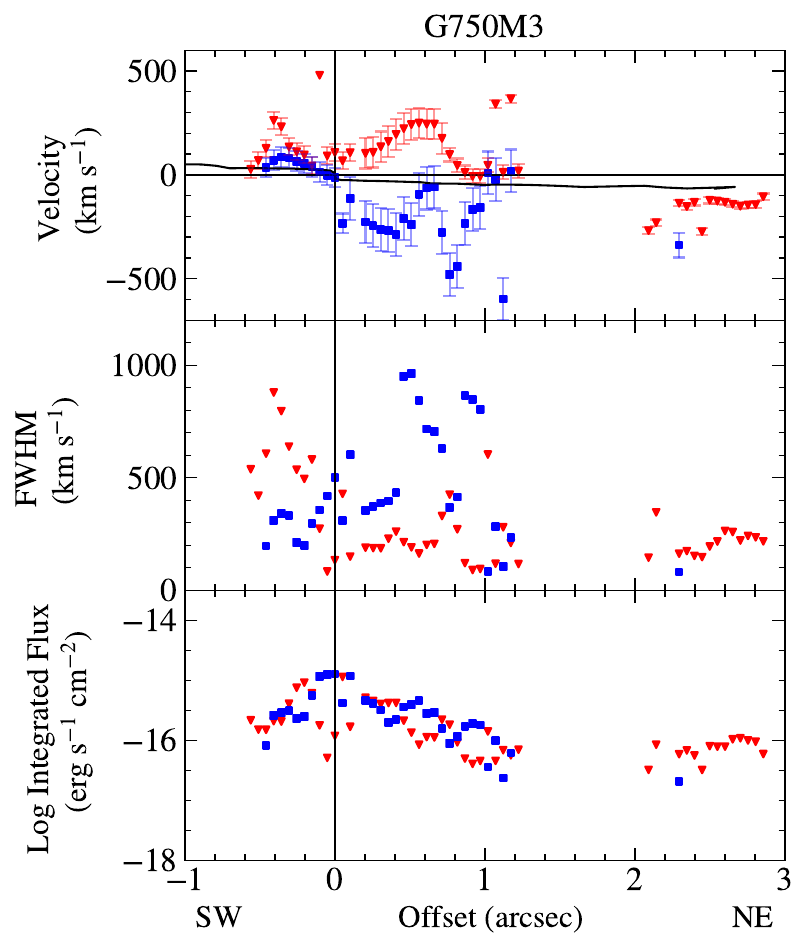}
\includegraphics[width=0.45\textwidth]{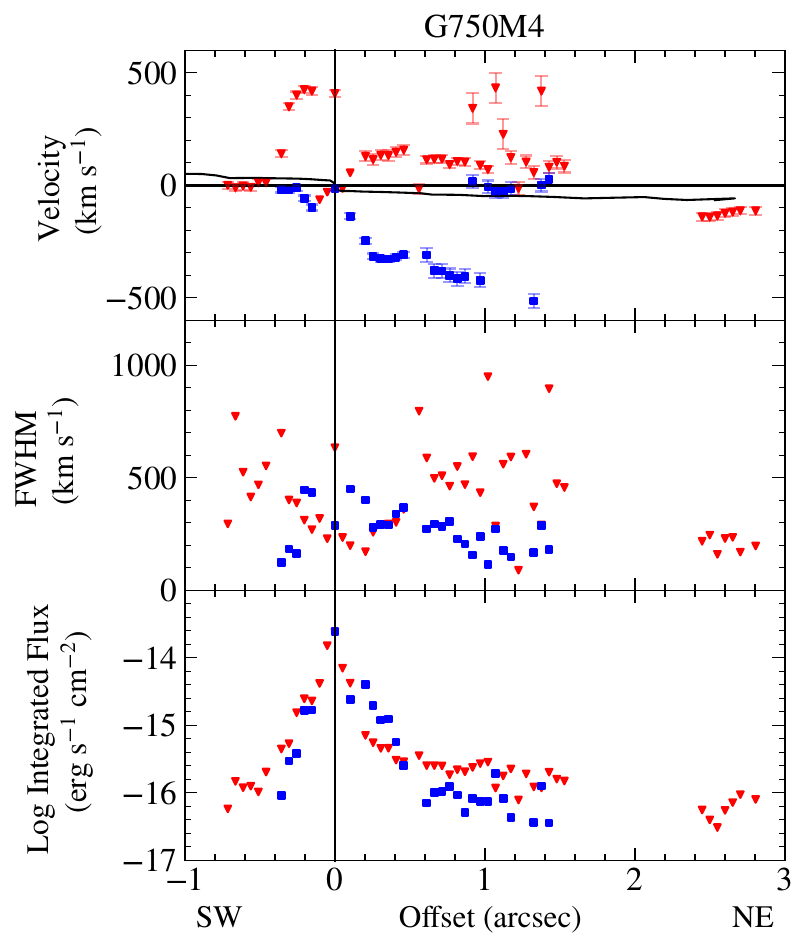}

\caption{The ionized gas kinematics of H$\alpha$ from the STIS G750M slits and [O~III] from the G430L slit. For slit position angles, refer to Table \ref{table:obs}. The top panel of each figure shows the radial velocity distribution of either [O~III]  (G430L) or H$\alpha$ (G750M) emission. The red triangles and blue squares represent the two kinematic components in order of decreasing velocity, respectively. Errors in velocity were calculated by taking the mean absolute deviation for distinct kinematic groups within each slit. The black curve is the rotation curve of the disk given by \citet{riffel17}. The middle panel shows the FWHM distribution of each component, and the bottom panel shows the integrated flux distribution. }
\label{fig: STIS kinematics}
\end{figure*}

\addtocounter{figure}{-1}
\begin{figure*}[]
 \centering  

\includegraphics[width=0.45\textwidth]{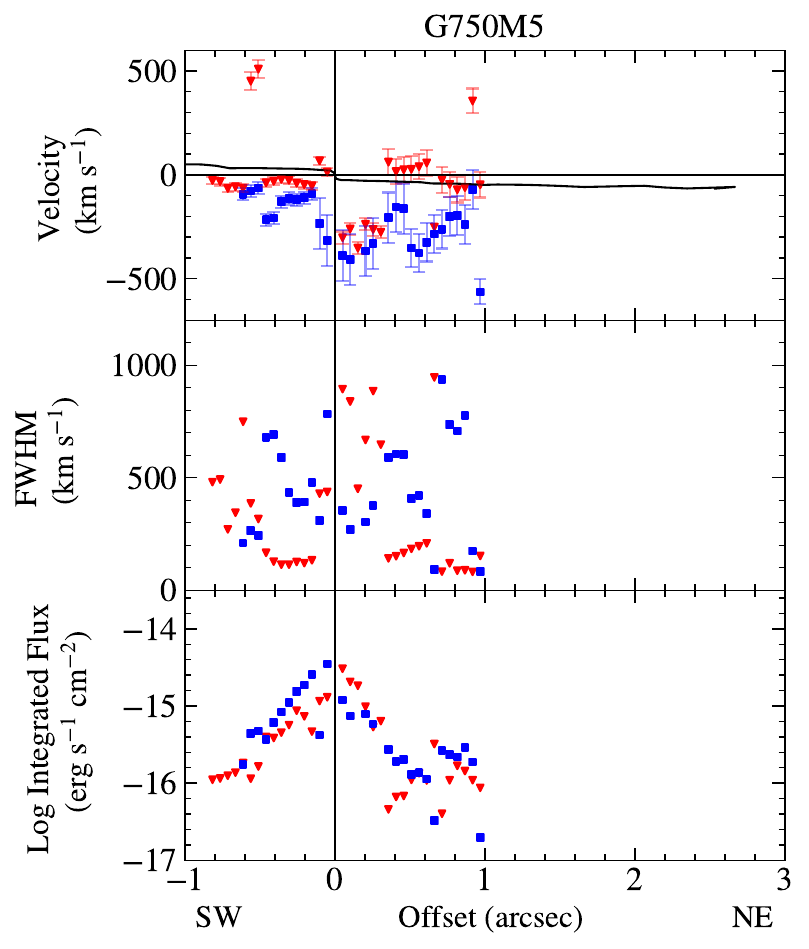}
\includegraphics[width=0.45\textwidth]{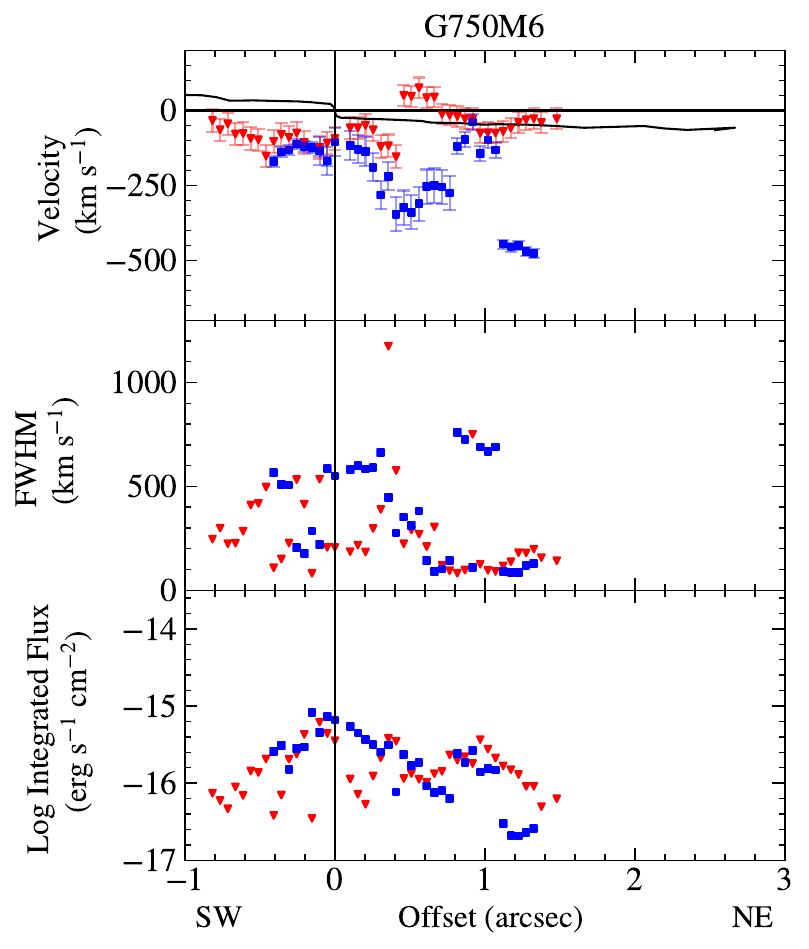}
\vspace{.25cm}
\includegraphics[width=0.45\textwidth]{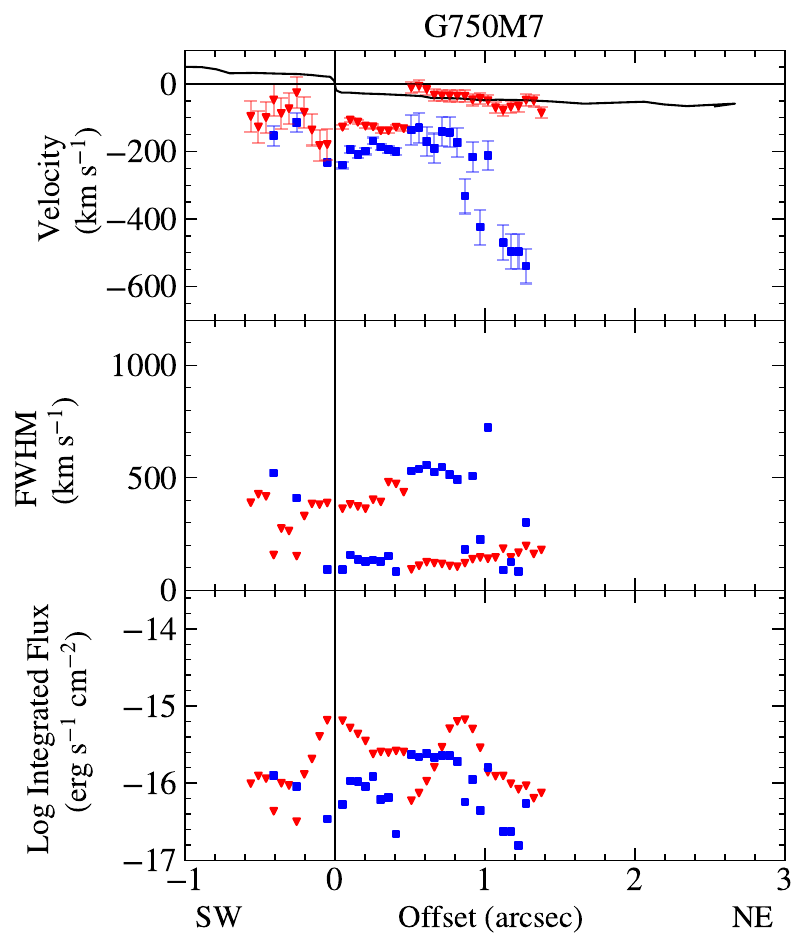}
\includegraphics[width=0.45\textwidth]{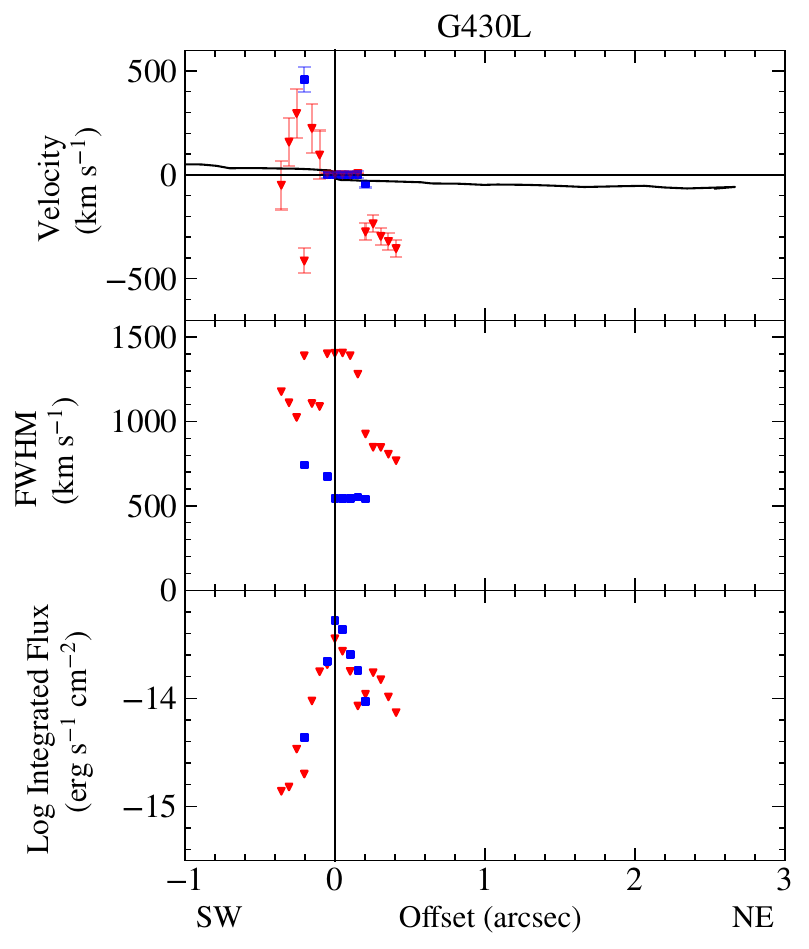}
\caption{Continued.}
\end{figure*}

\begin{figure}[t]
\centering  
\includegraphics[width=\columnwidth]{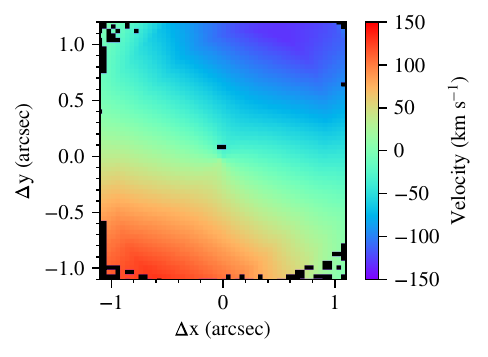}
\caption{The stellar velocity model for NGC 3227, taken from \cite{riffel17}. This LOS velocity field was created from NIFS K-band observations of the stellar CO absorption bandheads. We utilize this model in Section \ref{sec: nifs kinematics} to subtract the rotational motion from the observed velocity fields.} 
\label{fig: Rogemar model}
\end{figure}


\begin{figure*}[t]
\centering  

\includegraphics[width=\linewidth]{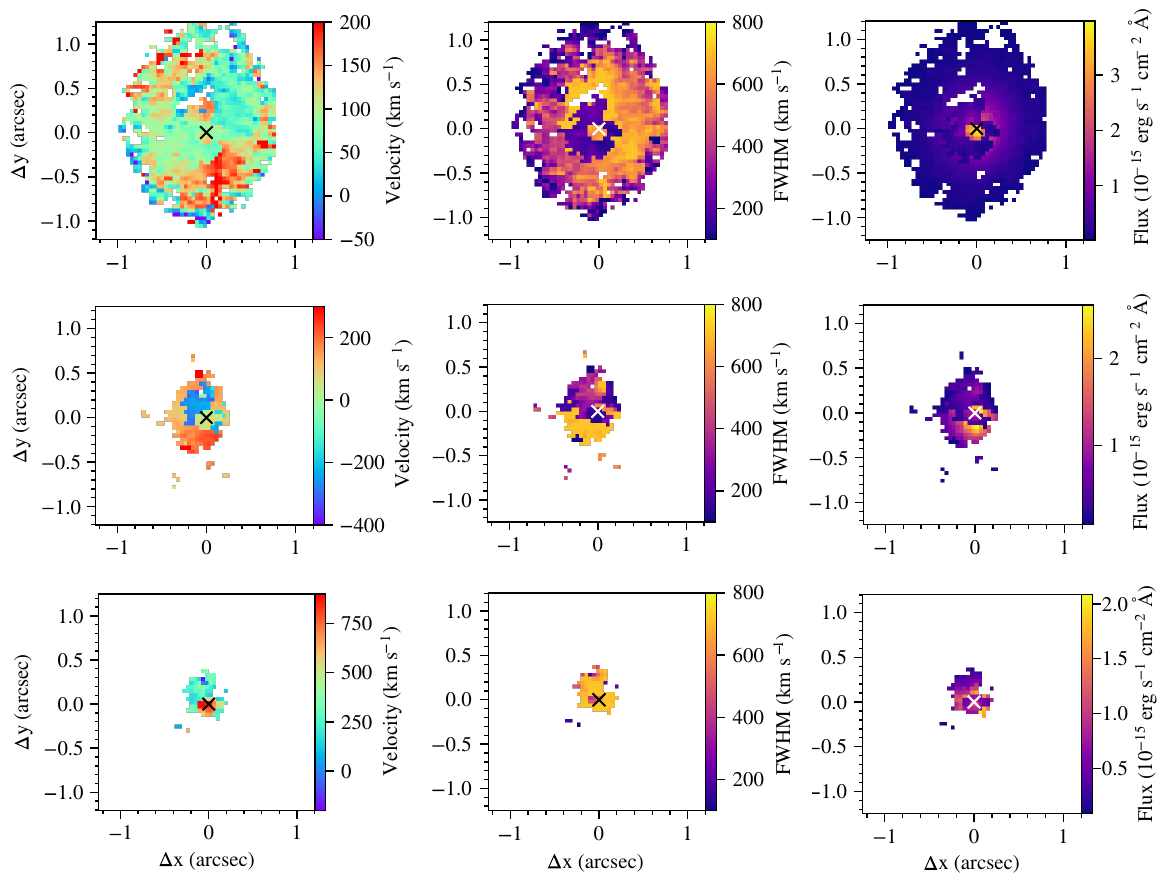}
\caption{Ionized gas kinematic plots for the velocity (left), FWHM (center), and flux (right) of the He I at $\lambda1.083~\mu$m emission from the NIFS Z-band. We incorporate the stellar kinematics model from \cite{riffel17} to subtract out the rotational velocities. The X at the center of each plot shows the position of the AGN. The top row shows the plots for the first component of each fit, the middle row shows the plots for the second component of each fit, and the bottom row shows the plots for the third component of each fit. The components are sorted by speed from lowest (top) to highest (bottom) at each position, with the bottom panel showcasing the high-velocity outflows.}
\label{fig: He I kinematics}
\end{figure*}

\begin{figure*}[p]
\centering  
\includegraphics[width=0.9\linewidth]{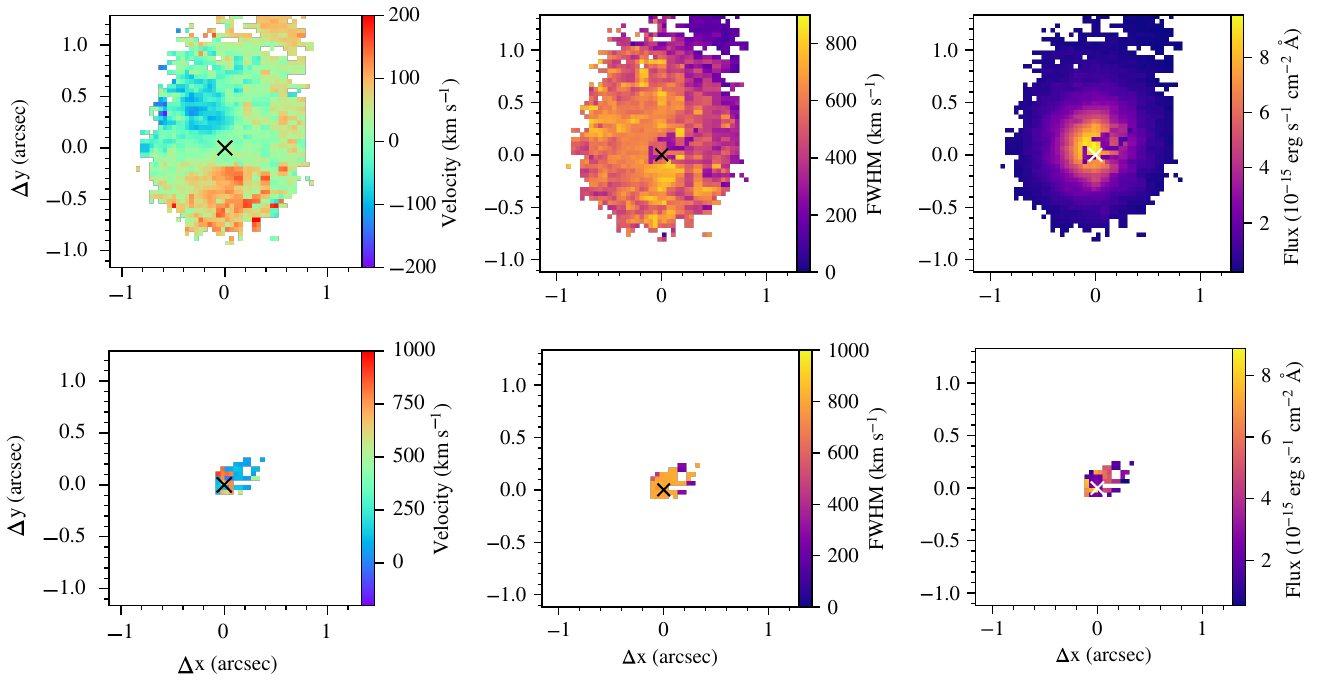}

\caption{Same as Figure \ref{fig: He I kinematics} but for [S~III] $\lambda9593$~\AA~emission from the NIFS Z-band.}
\label{fig: SIII kinematics}
\end{figure*}

\begin{figure*}[h!]
\centering  
\includegraphics[width=0.9\linewidth]{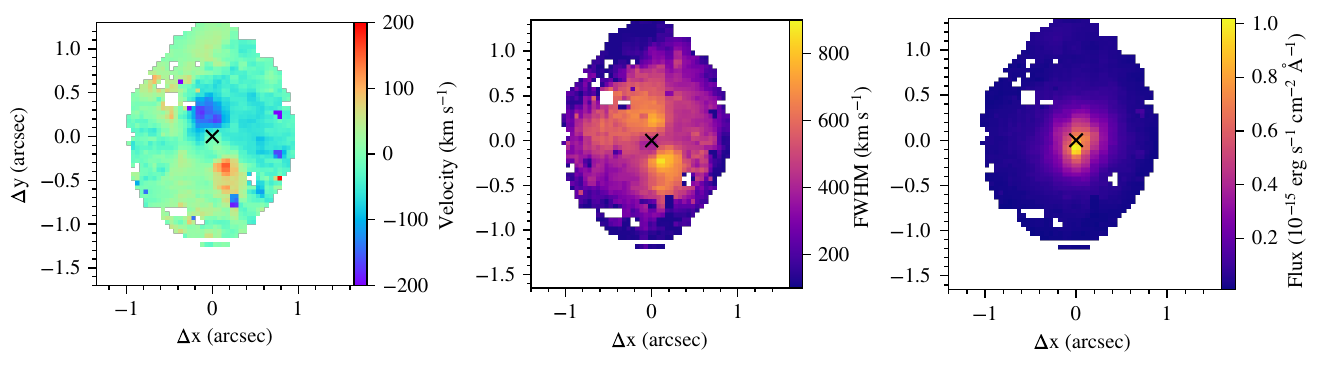}
\caption{Same as Figure \ref{fig: He I kinematics} but for Pa$\beta$ $\lambda1.2822~\mu$m emission from the NIFS J-band.}
\label{fig: PaB kinematics}
\end{figure*}

\begin{figure*}[!]
\centering  

\subfigure{\includegraphics[width=0.32\linewidth]{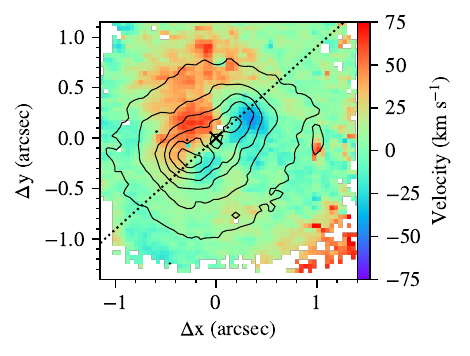}\label{fig: H2 vel}}
\subfigure{\includegraphics[width=0.32\linewidth]{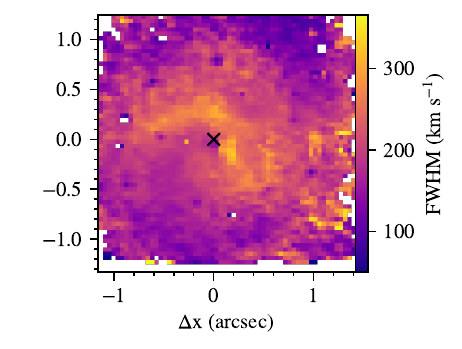}\label{fig: H2 FWHM}}
\subfigure{\includegraphics[width=0.31\linewidth]{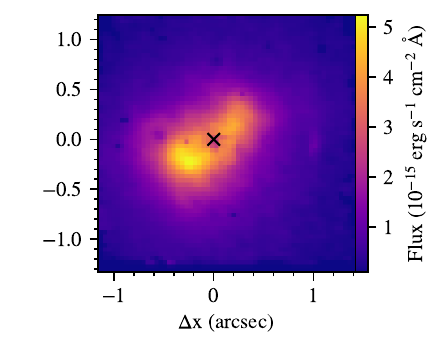}\label{fig: H2 flux}}

\caption{Same as Figure \ref{fig: He I kinematics} but for H$_2$ $\lambda2.1218$~$\mu$m emission from the NIFS K-band. In the leftmost plot, the velocity field is overlaid with the contour of the H$_2$ flux distribution shown in the rightmost plot. The dashed line in the velocity plot represents the major axis of the galaxy.}
\label{fig: H2 kinematics}
 \end{figure*}

Figure~\ref{fig:nucleus} shows how the STIS slits are generally oriented along the NLR axis, although some of the more extended outflows visible $4''$ to the north in the WFC3 image were not captured. The center of the G750M4 slit is located directly at the Seyfert 1 nucleus. Additionally, the STIS slits are close to the galaxy's minor axis, which minimizes the projected radial velocity due to rotation. The diminished impact of the rotational velocities is shown by the projected rotation curves in each plot of Figure \ref{fig: STIS kinematics}. For the STIS slits, we used the rotation curve from \cite{riffel17} rather than \cite{schinnerer00} because the former possesses a higher resolution needed for the small-scale STIS observations, although we still projected the rotation curve in the same manner described in Section \ref{sec: kosmos kin}. Assuming that outflows and rotation are the primary components comprising the kinematics of the circumnuclear region, we  infer that we are primarily seeing outflows in Figure \ref{fig: STIS kinematics} out to distances of 1\farcs3 (150 pc), with velocities reaching up to 500 km s$^{-1}$, similar to the largest amplitudes seen in the KOSMOS spectra. Across all G750M slits, the outflows appear to extend out to about 1\farcs5 from the center of the slit, or about 170 pc. However, this distance increases if we factor in the lateral distances from the slits, which are 0\farcs2 wide and separated by 0\farcs05, to the SMBH. When the slit locations are properly considered, we see outflows of 500 km s$^{-1}$ reaching 1\farcs7, or 200 pc, from the SMBH. Predominantly large FWHM values of the ionized gas out to the same distances are consistent with outflows. Unsurprisingly, the much higher spatial resolution of HST compared to APO allows us to better isolate the high-contrast outflowing knots of emission in the nuclear regions. The few points at larger distances ($\sim$3$''$) in the STIS data are consistent with rotation, although we don't detect the outflow component seen in Figure \ref{fig: KOSMOS major axis} at $\sim$5\arcsec, where there is spatial overlap between the KOSMOS and STIS data. This is likely due to the much smaller areas sampled by STIS, which is therefore unable to detect this component.

The [O~III] emission shown in the G430L slit reveals redshifted outflows in the of up to 500 km s$^{-1}$ located 0\farcs2 SW of the nucleus. We also see blueshifted outflows on the order of 300-400 km s$^{-1}$ out to 0\farcs4 NE. 

We also see velocity trends across several slits. For instance, many slit positions show similarly blueshifted and redshifted velocities, indicating the presence of a bicone that is funneling gas both towards and away from our line of sight along the diametrically positioned cones that form the bicone structure. Additionally, a curious feature across several slits is the presence of blueshifted velocities peaking at $-$500 km s$^{-1}$ that are 0\farcs5 -- 0\farcs8 NE of the nucleus. We speculate on the source of this outflow in the Section~\ref{sec: discussion}.


\subsection{Gemini NIFS Kinematics}
\label{sec: nifs kinematics}

Kinematic maps, created from our reduced NIFS J-, K-, and Z-band data, display the velocity, FWHM, and flux distribution for up to three components in Figures~ \ref{fig: He I kinematics}--\ref{fig: H2 kinematics}. We have subtracted the rotational motion from the velocity maps using the stellar velocity model for NGC 3227 given in \cite{riffel17} and shown in Figure \ref{fig: Rogemar model}, so it can be assumed that mapped velocities are due to non-rotational (i.e. outflowing) motions.

Our velocity maps therefore differ somewhat from those previously published using these and other IFU observations of the nuclear region in NGC~3227 \citep{davies06, barbosa09, schonell19, bianchin22}, which show the \textit{observed} velocity fields containing the composite rotational and outflowing motions, often using only single-component fits.

\subsubsection{Z-band Kinematics}
Figure~\ref{fig: He I kinematics} shows the three-component fits for the He~I $\lambda1.0830~\mu$m emission in the Z-band. We fit both broad and narrow components, but will focus on the latter in this analysis. The bottom panel shows the presence of an extremely concentrated but high-velocity ($\sim$1000 km s$^{-1}$) outflow in a redshifted knot just SW of the nucleus.
The middle panel shows a lower-amplitude and more extended pattern of outflow that is primarily blueshifted to the NE and redshifted to the SW. The upper panel shows an extended component that can be attributed to rotation (after subtraction of the rotation model), except for perhaps a mildly redshifted component to the SW once again. 


Figure~\ref{fig: SIII kinematics} shows the two-component fits for the [S~III] $\lambda9533$ emission in the Z-band. The components are sorted in order of decreasing velocity: the bottom panel highlights the high-velocity outflows and the top panel exhibits the lower-velocity outflow structure. Along the top panels, we see redshifted outflows on the order of 100 -- 150 km s$^{-1}$ extending southwards over a distance of almost 1$''$ from the nucleus, with some areas reaching speeds of almost 200 km s$^{-1}$. We also see a blueshifted outflow extending towards the NE direction from the nucleus, with speeds of $-$150 to $-$100  km s$^{-1}$ out to an orthogonal distance of 0\farcs75. The FWHM plot for this map does not have many discernible features, although the values are notably smaller on the west end than the east end by about 200 km s$^{-1}$. The flux distribution is relatively smooth compared to the HST [O~III] image, because the Gemini NIFS PSF has much broader wings than that of HST and its cameras.

\begin{figure*}[t]
\centering  
\subfigure{\includegraphics[width=0.45\linewidth]{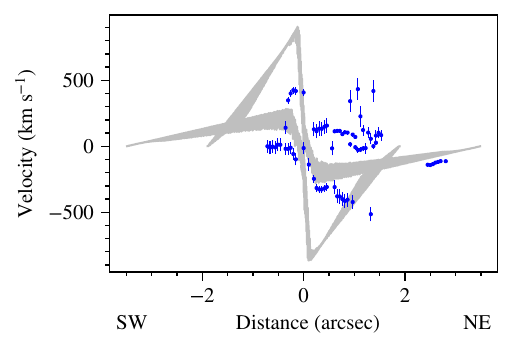}}
\subfigure{\includegraphics[width=0.45\linewidth]{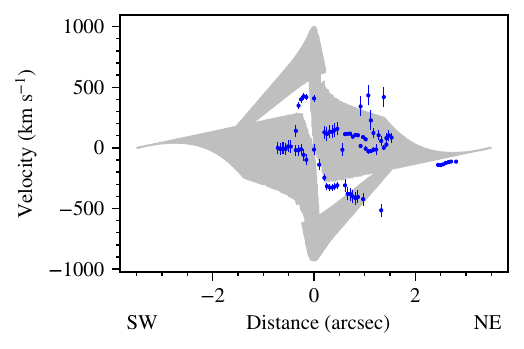}}
 \caption{A demonstration of how the bicone model changes based on whether one chooses the range of bicone opening angles to be small (7\arcdeg, left) or large (39\arcdeg, right). All other bicone parameters are the same. The gray areas represent the model velocity space, and the blue points are data from the corresponding STIS slit that we aim to match with the models.}
\label{fig: bicone opening angles}
\end{figure*}

The lower set of panels in Figure~\ref{fig: SIII kinematics} shows high-velocity outflows in the inner 0\farcs25 region around the nucleus, with speeds reaching up to 900 km s$^{-1}$. The chaotic motion often associated with outflows is also reflected in the corresponding FWHM plot, which shows line widths as large as 800 km s$^{-1}$. 


\subsubsection{J-band Kinematics}
Figure~\ref{fig: PaB kinematics} shows the emission of Pa$\beta$ $\lambda1.2822~\mu$m seen in the J-band. Unlike the lines observed in the Z-band, only a single component was needed to sufficiently fit Pa$\beta$. The velocity plot shows clearly defined blueshifted and redshifted lobes with their centers $\sim$0\farcs3 from the nucleus in the NE and SW directions, respectively. Stretching outwards from the west side of the blueshifted lobe is another region of blueshifted outflow, albeit with a much lower $|v_r|$ of only $\sim$80 km s$^{-1}$. We do not see a similar trend in the redshifted lobe. We see a correspondence between the velocity and FWHM plots, as the regions with high $|v_r|$ show elevated line widths reaching 850 km s$^{-1}$. Curiously, we see very low line widths in the region with the secondary blueshifted outflow. The flux profile in Figure~\ref{fig: PaB kinematics} shows a clear radial decrease in flux from the brightest point in the center, which corresponds to the SMBH.

Overall, the emission lines from ionized gas in the NIFS observations show high-amplitude blueshifted velocities in the NE and high-amplitude redshifted velocities in the SW, reaching maximum values of 400 -- 800 km s$^{-1}$ at distances 0.2$''$ -- 0.4$''$ from the SMBH. Lower amplitude blueshifts and redshifts that cannot be attributed to rotation can be seen on either side of the nucleus over the NIFS field of view. This pattern is very similar to that seen in the HST STIS observations.

\subsubsection{K-band Kinematics} \label{sec: H2 kin}
Figure~\ref{fig: H2 kinematics} shows the emission of H$_2$ $\lambda2.1218$~$\mu$m seen in the K-band. We see interesting structure in all three of the kinematic plots from our single-component fit, also seen but not as pronounced in the observed velocity fields by \cite{schonell19} and \cite{bianchin22}, likely due to differences in fitting methods. The rotation-subtracted velocity map in Figure~\ref{fig: H2 vel} shows two adjacent redshifted and blueshifted lobes on the NW and NE sides of the nucleus, respectively. The red lobe expands northwards out to $1''$, while the blue lobe is more concentrated and has a width of $\sim$0\farcs4. The speeds that we see in the H$_2$ maps are much lower than those in the other NIFS velocity maps, with maximum values of $|v_r|\approx 70$ km s$^{-1}$, indicating that the H$_2$ kinematics are rotation-dominated. The dashed line marking the major axis of the host disk reveals the symmetry of the blueshifted and redshifted emission along it. The lower right corner shows an area that appears to also have redshifted emission, but due to the low FWHM and flux values, those values could be due to noise in the observed spectra. The FWHM plot in Figure~\ref{fig: H2 FWHM} reveals an unusual curved structure that coincides with regions of the redshifted emission.

In Figure~\ref{fig: H2 flux}, we see a rather unique bean-like shape about $1''$ long and centered over the SMBH. We also overlay the contours of the flux map onto Figure~\ref{fig: H2 vel} and observe that the redshifted emission corresponds to a region 2$-$3 times fainter than the peak emission. However, we also see that the blueshifted emission is cospatial with the flux contours on the northern end.

\section{Bicone Model}
\label{sec: bicone}

\cite{travisthesis} describe the structure of a biconical outflow model for NGC 3227's NLR. They utilized the kinematic modeling code from \cite{das05} to simulate models dependent on input parameters including the bicone's orientation on the sky, inclination with respect to our line of sight (LOS), minimum and maximum half-opening angle (HOA), turnover radius, and height. The model also follows a velocity profile which starts at 0 km s$^{-1}$ in the center and increases linearly until a specified turnover radius, after which the velocity decreases linearly until zero at a predetermined maximum height along the bicone axis. This empirical velocity law appears to match the velocity trend for many Seyfert NLRs quite well \citep{travisthesis}.

For a given set of input parameters, the model produces a plot that shows the range in possible velocities as a function of distance from the AGN. Figure \ref{fig: bicone opening angles} shows the ways in which these plots can vary by changing the range in HOA, but it also exemplifies a characteristic bowtie-like shape whose peaks reflect the chosen turnover radius.  For each STIS slit, \cite{travisthesis} compared the measured [O~III] velocities to the velocity range given by the model and adjusted the input parameters accordingly to find the best match. In the absence of a practical statistical algorithm, the best match between the model and data was determined by eye. However, Figure D.7 of \cite{travisthesis} shows large discrepancies between the model and data for NGC~3227. This paper defines a new biconical outflow model that provides a better match for the data. 


\subsection{Development of a New Model} \label{newmodel}
Our goal is to determine a bicone model that most concisely fits the data. We used the same code utilized in \cite{travisthesis} and improved the selection of the model that best fits the data. 

In order to determine the optimal model, we used the bicone parameters given by \cite{travisthesis} as priors and generated over 300,000 bicone models with corresponding velocity plots like those shown in Figure \ref{fig: bicone opening angles}. One condition we implemented was that the bicone inclination angle from the LOS must have a smaller value than the inner opening angle, because the Seyfert 1 designation for NGC~3227 requires visibility of the central engine. We iterated through five variable parameters (bicone inclination, inner opening angle, outer opening angle, turnover radius, and bicone PA), creating a model for every combination of parameters within $\pm $10\arcdeg of prior estimates for each STIS slit position, using step sizes of 1\arcdeg.  Two parameters that we did not vary are the maximum height and maximum velocity, which we kept constant at 150 pc and 600 km s$^{-1}$, respectively. We chose these values based on the kinematics of our data, and although these values are slightly different than those reported in \cite{travisthesis}, the difference is negligable. 

Each model shows the projected bicone kinematics at each of our STIS slit positions. We refer to the ``model velocity space'' as the gray regions in Figure \ref{fig: bicone opening angles}.
For each slit, the algorithm determines the ``velocity fraction,'' which is the fraction of data points that fall within the model velocity space and characterizes how well the model fits the data. However, our data represent a mix of rotating and outflowing motions, and because we want to use outflows to determine the orientation of the NLR bicone, we removed data exhibiting characteristics of rotational motion. \cite{riffel17} defined a rotation curve for the inner 2$''$ of NGC 3227, and in the absence of uncertainties from that model, we assume that data that fall within $\pm$20 km s$^{-1}$ of the rotation curve and outside the model velocity space are rotational. We subtract the number of rotational data points from the total number of points that we fit with the model to ensure that we are only fitting outflowing data to the bicone model. 
\begin{table}
  \begin{tabular}{|c|c|c|} \hline
  Parameters & Fischer et al. & This Work \\ \hline
  Galaxy PA & $-31$\arcdeg&  $-31$\arcdeg\\
  Galaxy Inclination &63\arcdeg (SW closer)& 48\arcdeg (SW closer)\\ 
  Bicone PA & 30\arcdeg& 27 $^{+4}_{-2}$ \arcdeg\\
  Bicone Inclination& 15\arcdeg$^\dagger$ (SW)& 40$^{+5}_{-4}$\arcdeg (SW)\\
  Inner HOA& 40\arcdeg& 47$^{+6}_{-2}$\arcdeg\\
  Outer HOA& 55\arcdeg& 68$^{+1}_{-1}$\arcdeg\\
  Turnover radius & 100 pc& 26$^{+6}_{-6}$ pc\\
  Max height & 200 pc& 150 pc\\
  Max velocity & 500 km s$^{-1}$& 600 km s$^{-1}$\\
  
  \hline
  \end{tabular}
   \caption{Rows are (1) major axis of galaxy from \cite{schmitt00}, (2) inclination of galaxy from \cite{xilouris02}, (3) major axis of bicone axis, (4) inclination of bicone axis relative to plane of sky, (5) minimum HOA relative to the LOS, (6) maximum half opening angle, (7) turnover radius, (8) maximum height of bicone along its axis, and (9) maximum outflow velocity, located at turnover radius. Errors for this work are reported as a 68\% confidence interval.  \\\hspace{\textwidth}  
   $^\dagger$ \cite{travisthesis} lists inclination from the plane of the sky, whereas we opt to use the angle from our LOS. Their value has been converted to our frame of reference for consistency.}
 \label{table: bicone params}
  \end{table}

The parameters that most drastically impact the geometry are bicone inclination, the inner and outer half-opening angles of the bicone, the turnover radius, and the bicone position angle. The opening angles had to be considered carefully because they can easily lead to model overfitting. Figure \ref{fig: bicone opening angles} shows the effect that a small and large range of opening angles can have on the resulting bicone model. A bicone with a wide range in the opening and closing angles will always envelop a larger fraction of the data than a bicone with a narrower angular range. Simply determining the model with the highest velocity fraction is insufficient to finding the optimal bicone model. Thus, we must apply a normalizing weight based on the size of the model velocity area to act as a counterbalance. For each slit position, we compiled all 2D model velocity areas into histograms, which revealed normal distributions. We use those distributions to weight each model velocity area, and divide the velocity fraction by the weighted model velocity area to determine the density of fitted points per unit model velocity space, which we call the ``weighted velocity fraction.''

\begin{figure}[t]
  \includegraphics[width=0.95\linewidth]{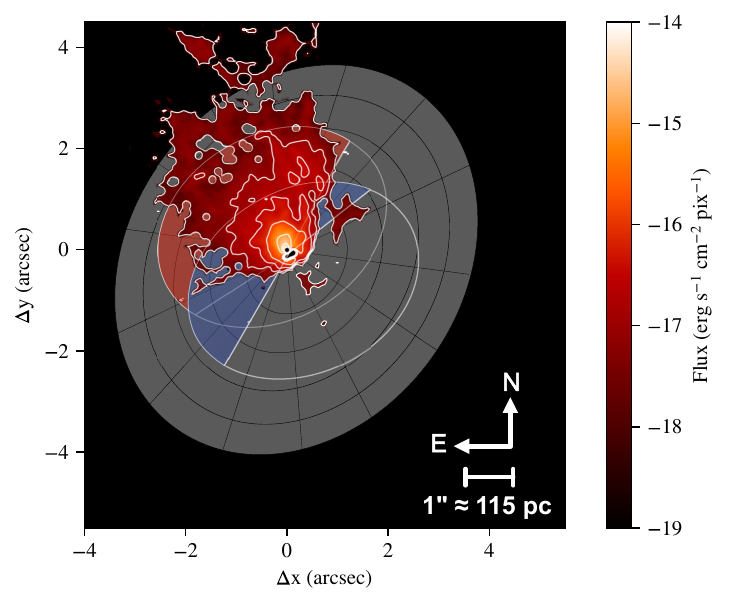}%
{%
  \caption{Projection of our new bicone model, outlined in white, overlaid on the WFC3 [O~III] image, shown by the contours. The gray disk represents the plane of the galactic disk, which intersects with the [O~III] line emission. The unobscured part of the near cone is shaded in blue, while the unobscured part of the far cone is shaded in red. In this orientation, the [O~III] emission is entirely in front of the bicone and is blueshifted.}%
  \label{fig: bicone on WFC3 image}
}


\end{figure}

We average the weighted velocity fractions for all seven STIS slits corresponding to a given set of input parameters and deemed the set of parameters with the highest average to be the optimal model, for it maximizes the number of fit data points per unit area. Those parameters are given in Table \ref{table: bicone params}, with uncertainties marking parameter values that reside within the 68\% confidence interval for the average velocity fractions.

\begin{figure*}[t]
\centering 

\subfigure{\includegraphics[width=0.9\linewidth]{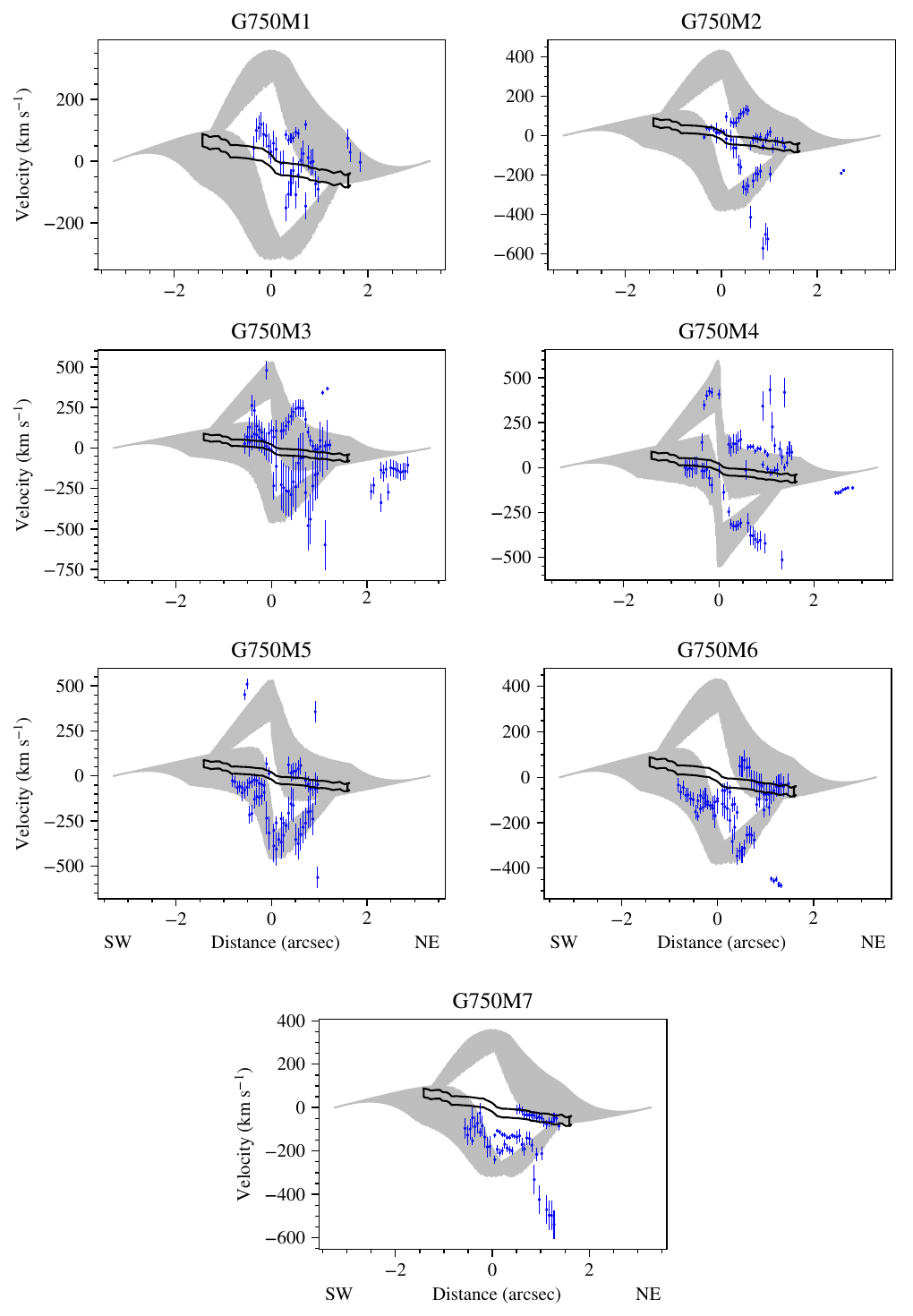}}




\vspace{-.5cm}

\caption{Each figure shows the kinematic profile of a STIS slit. The blue points show the observed [O~III] $\lambda$5007 velocities. The gray regions show the model velocity areas, and represent the range of acceptable velocities for the updated version of the bicone model. We define the area of stellar rotation, which is encompassed by the black outline, as points which fall within $\pm$ 20 km s$^{-1}$ of the rotation curve given by \cite{riffel17}.}
\label{fig: STIS velocity models}
\end{figure*}

Table \ref{table: bicone params} shows that there are significant changes in the new bicone model compared to that described in \cite{travisthesis}. Figure \ref{fig: bicone on WFC3 image} shows the projection of the bicone on to the same WFC3 image of [O III] emission as Figure \ref{fig:fig1}. Most notable from \cite{travisthesis} is the change in inclination, which steepened significantly from 15\arcdeg to 40\arcdeg. This value is in much better agreement with the measurement of $33.2^{+13.5}_{-9.1}$\arcdeg from \cite{bentz23}, based on modeling of the broad-line region (BLR), and suggests that the confinement structures for the BLR (accretion disk) and NLR (torus) are co-aligned. Another difference is in the turnover radius, which describes the spot where deceleration overtakes acceleration and is shown in projection as the locations of the maximum absolute velocities in Figure \ref{fig: STIS velocity models}. Whereas \cite{travisthesis} found the turnover radius to be 100 pc, we determine it to be 26$^{+6}_{-6}$ pc. Additionally, we find an outer HOA of 68$^{+1}_{-1}$\arcdeg, which also aligns with the value of 64.7$^{+18.3}_{-11.8}$\arcdeg determined in \cite{bentz23}. 

\subsection{Comparison of Data to Model}
Figure~\ref{fig: STIS velocity models} shows how the range of projected velocities from the model compares to the velocities from the STIS data. We can divide the plot into four quadrants. In the figure, the negative direction refers to the SW. The top half represents emission from the further redshifted cone, while the bottom half shows emission from the nearer blueshifted cone. The disk of the galaxy intersects with the AGN in such a way that positive distances show emission primarily from the part of the bicone in front of the disk, while negative distances show emission from the part of the bicone behind the disk. There is not much emission from the SW in any of the slits due to the prominent dust lanes from the disk that block our view (see Figure \ref{fig: Judy Schmidt}), except within the inner $\sim$0\farcs8 where the [O~III] emission is bright enough to be seen through the extinction.


The overlap between the model and data in this work is substantially improved from that in \cite{travisthesis}, and we see generally decent agreement between data and model in all of the slits. In a number of slits, we see many points that follow the rotation curve, indicating strong rotational motions in the inner 1$''$, which is already known \citep{riffel17, bianchin22}. This rotation is also seen in the data in slits G750M2, G750M3, and G750M4 in the regions beyond 2$''$, as shown in Figure~\ref{fig: STIS kinematics}. 

There are two main discrepancies between the models and observed velocities. First, the outside slits (G750M1, G750M2, G750M6, G750M7) show observed velocities in the inner regions that don't quite match either rotation or outflows, particularly on the NE side. 
These may be due to disturbed kinematics \citep{fischer18}, where outflows are running into the dust lanes, consistent with the large FWHM values for most of these points \citep{fischer19}.
Second, there is the large clump of blueshifted velocities peaking at $-$500 km s$^{-1}$ that is 0\farcs5 -- 0\farcs8 NE of the nucleus in most of the slits, which was previously mentioned in Section \ref{sec: STIS kinematics}. In Figure \ref{fig: STIS velocity models}, we see that a number of those points lie outside the model. We discuss this discrepancy in Section \ref{sec: discussion}.

\section{Results and Discussion}
\label{sec: discussion}

\subsection{Connection to NGC 3227's Circumnuclear Ring}
\label{circumnuclear ring}

Using observations of CO and HCN with the IRAM Plateau de Bure interferometer (PdBI) at 0\farcs6 resolution, \cite{schinnerer00} found a 3$''$ (345 pc) diameter molecular ring of gas centered near the SMBH with much stronger emission on its eastern side. The ring is at PA = 160\arcdeg\ and inclination = 56\arcdeg, in agreement (to within $\sim$10\arcdeg) with the large-scale H~I 21 cm \citep{mundell95}, large-scale stellar \citep{fischer17}, and nuclear stellar \citep{riffel17} disks. Between 1$''$ and 3$''$, the ring rotates in the same way as the large-scale gas and stars, but the emission appears to show either counterrotation or noncircular motions inside of 1$''$.
\cite{schinnerer00} found that the kinematics of the molecular gas couldn't be replicated by a bar in the inner 1$''$ of NGC 3227, but instead proposed a molecular ring that starts 75 pc from the nucleus and bends to be perpendicular to the host galaxy plane at a distance of 30 pc. The nuclear kinematics in both cold and warm molecular gas have been a topic of study for over a decade \citep{alonso19, davies14, hicks08, schonell19}. 

\cite{alonso19} present ALMA CO observations of the nuclear molecular ring at 0\farcs1 -- 0\farcs2 resolution and confirm strong noncircular motions at distances $\leq$ 1$''$ from the nucleus, primarily along the disk's kinematic minor axis. They find that the noncircular motions can best be explained by {\it outflow} from the inner part of the ring that lies along the PA and inclination of the host disk, with the far side of the ring in the NE direction. They get a slightly better fit by including a warp with the far side of the inner ring instead in the SW direction, as suggested by \cite{schinnerer00}. However, this would imply {\it inflow} at deprojected velocities of at least 150 km s$^{-1}$ assuming their interpretation of disk which is substantially higher than the inflow velocities expected of lower luminosity AGN, which is on the order of $20-50$ km s$^{-1}$ \citep{davies06}. They also demonstrate that infrared IFU observations of H$_2$ from \cite{davies14} show a similar velocity field for the warm molecular gas.


Our NIFS H$_2$ radial velocity, dispersion, and flux maps shown in Figure \ref{fig: H2 kinematics}  are similar to those from other studies \citep{riffel17, alonso19, davies14, hicks08, bianchin22}. However, the rotation-subtracted velocity map shows the axis of the presumed H$_2$ outflow slightly offset from the nucleus and at a PA $\approx$ 90\arcdeg, compared to the cold ring's minor axis at 50\arcdeg -- 70\arcdeg \citep{schinnerer00, alonso19}. Moreover, redshifted emission at $\sim$60 km s$^{-1}$ is visible almost 1$''$ from the nucleus, while the blueshifted outflows are much more concentrated. 
Due to the high deprojected velocities in both the warm and cold molecular gas (up to 150 km s$^{-1}$) and their similar orientations with respect to the large-scale disk, we prefer the interpretation of redshifted emission in the east and blueshifted emission in the west as molecular outflow.

With this interpretation, the geometry of our biconical outflow model and galactic disk as shown in Figure \ref{fig: bicone on WFC3 image} provides a natural explanation for the observed kinematics of both ionized gas and cold/warm molecular gas. For the ionized gas, the NE side of the near bicone is unobscured, has high projected blueshifts, and therefore dominates the [O~III] images and optical spectra. The unobscured portion of the redshifted NE side of the far cone also contributes significantly, especially to the more extended emission. The molecular outflows originate in the intersection between the bicone and host galaxy disk and move along the disk. The longer-wavelength observations experience smaller or no extinction compared to the optical, so that we see redshifts in the east and blueshifts in the west. 
Our observations do not rule out a warp in the disk inside of $\sim$30 pc, and in fact there is likely to be an interior warp to provide a torus-like structure (at PA $\approx$ $-$63\arcdeg, inclination $\approx$ 27\arcdeg, NE side closer) that produces the observed bicone. 



\begin{figure}[t]
\centering  
\includegraphics[width=\linewidth]{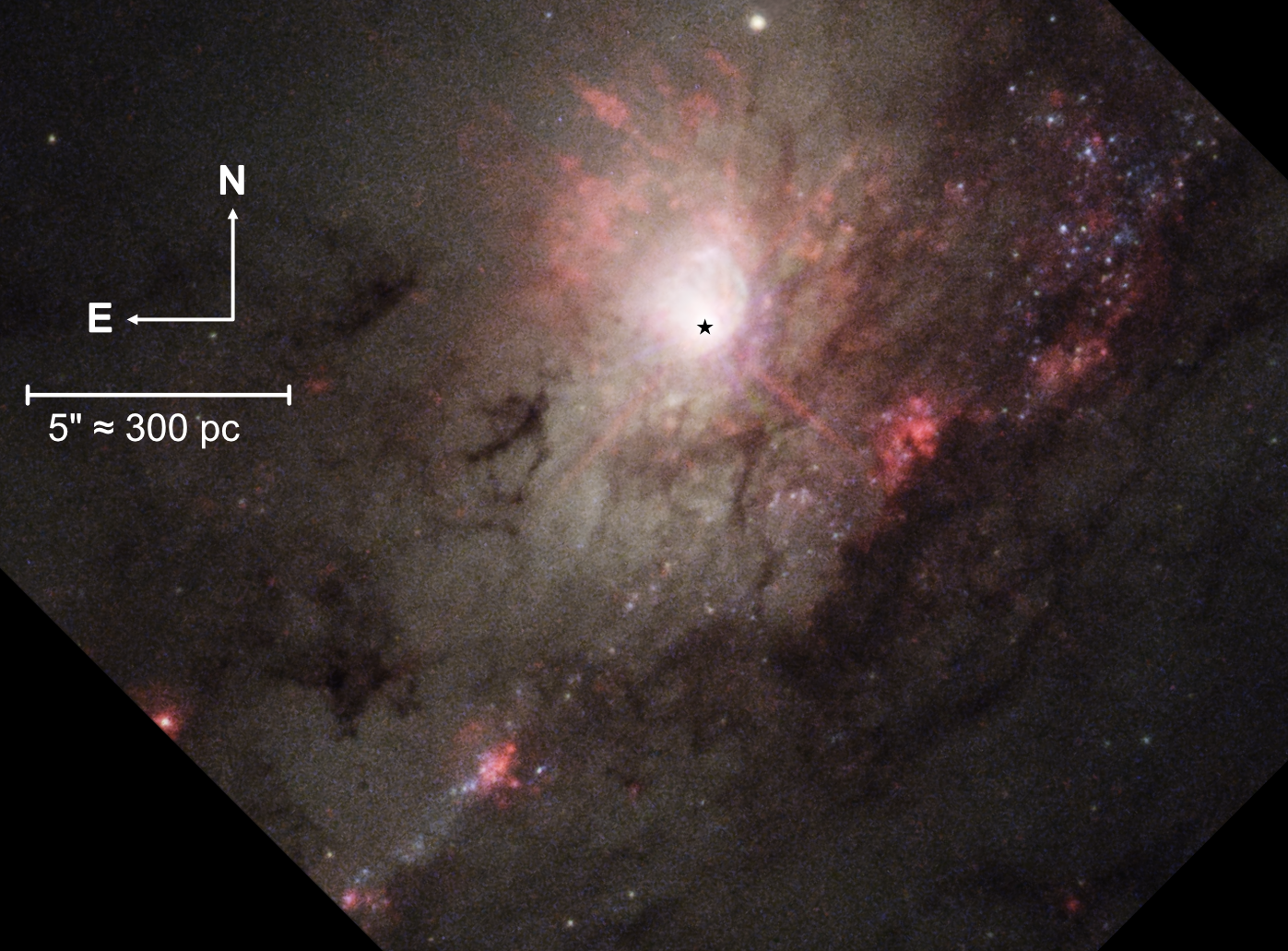}
\caption{Color-composite image of the inner region surrounding NGC 3227’s AGN where red colors are from the HST WFC F658N filter, green colors are from the HST WFC3/UVIS F547M filter, and blue colors show F550M and F330W filters from Hubble’s Advanced Camera for Surveys High Resolution Channel (HRC). The black star shows the location of the nucleus. Image credit: Judy Schmidt.} 
\label{fig: Judy Schmidt}
\end{figure}

\subsection{Kinematic Properties of the Ionized Gas}

\cite{fischer18} categorizes the observed motions of the ionized gas in the NLR into three groups:
\begin{enumerate}
\item Motion indicative of outflowing material. These data points are typically characterized by a high velocity ($\geq$ 400 km s$^{-1}$) and FWHM ($\geq$ 300 km s$^{-1}$), and are representative of a source of acceleration needed to account for velocities too large for gravitational acceleration alone.
\item Motion caused by the rotation of the galactic disk. These points are typically characterized by a low velocity ($\leq$ 400 km s$^{-1}$) and FWHM ($\leq$ 300 km s$^{-1}$), and show a symmetric pattern of blueshift/redshift on either side of the nucleus. 
\item Motion affected by a kinematic ``disturbance." This is gas affected by an impact or turbulence that could cause a redistribution of the direction of motion. The disturbed gas shows low velocities ($\leq$ 400 km s$^{-1}$) and high FWHM ($\geq$ 300 km s$^{-1}$), as the gas is being accelerated but its motion is impeded to some extent by its surroundings.
\end{enumerate}

Separating the observed kinematics into these three groups can be challenging. For example, rotational motion and outflowing motion can have similar kinematic signatures if the outflow is in the same direction as the rotation. Another possibility is that a measurement of low velocity may actually be high-speed outflowing material whose motion is primarily across our line of sight. Nevertheless, by comparing the observed large-scale velocities to the rotation curves from the stars or cold gas as discussed in Section \ref{kinematics}, we can separate these motions with a good deal of confidence.

\subsubsection{Extents of the Outflows and Disturbed Gas}

In Figure \ref{fig: KOSMOS major axis}, H$\alpha$ emission shows consistently sustained outflows of up to $-500$ km s$^{-1}$ out to $5''$ ($\sim$600 pc) in the faintest of three components.
This extent is significantly greater than the height of the bicone model ($\sim$200 pc), which we determined based on the STIS data. However, the bicone height does not act as an upper limit to the extent of the outflows, but rather correlates with the turnover radius. Nevertheless, these high velocity points are still within the bicone and consistent with radiative driving from the innermost parsecs, as shown in Section \ref{radiativedriving}.
Additionally, in Figure \ref{fig: KOSMOS major axis}, the velocities from 3$-$7$''$ SE are practically 0 km s$^{-1}$, which are substantially lower than the velocities predicted by the rotation curve in that region of around 200 km s$^{-1}$.
This is likely due to averaging of two components (redshifted rotation, blueshifted outflow) that could not be separated due to their large FWHM.
At distances $< |15|''$ (1.7 kpc), there is a clear rotational component (primarily brighter points) extending to the nucleus, although most of this component appears to be ``disturbed" with large FWHM.
Points at larger distances (up to 75$''$ or 8.6 kpc) are clearly due to pure rotation.
%

The projection of rotational kinematics is minimized along the minor axis, so the observed motions in Figure~\ref{fig: KOSMOS minor axis} must be primarily due to outflows, albeit at projected velocities $\leq$ 250 km s$^{-1}$. Thus, it is puzzling to see $\sim$100 km s$^{-1}$ emission at $20 -30''$ SW, especially in the blueshifted direction when we expect any rotational motion to be redshifted.
The kinematics along the outflow axis (Figure \ref{fig: KOSMOS outflow axis}) show evidence for pure rotation at large distances ($\sim$55$''$), disturbed rotation in the higher velocity component from $-$10$''$ to $\sim$10$''$, and outflows in the lower velocity component out to $\sim$7$''$ ($\sim$800 pc).

Overall, our results indicate that the ionized outflows in NGC 3227 extend to distances of about 800 pc from the central engine. However, the majority of the ionized gas is kinematically disturbed out to $\sim$15$''$ (1.7 kpc), a large fraction of the bulge size, which would make it difficult to form stars from this gas.





A curious feature across several slits in Figure~\ref{fig: STIS kinematics} is the presence of blueshifted velocities peaking at $-$600 km s$^{-1}$ and positioned 0\farcs5 -- 1\farcs2 NE. In the slits G750M2, G750M3, G640M4, G750M5, G750M6, and G750M7, $|v_r|$ increases with distance, which is a clear sign of outflowing motion as described by the radiative driving formulism (see Section \ref{radiativedriving}). However, Figure~\ref{fig: STIS velocity models} shows that a number of data associated with those outflows fall outside the model. Gas clouds that have been accelerated to high speeds at significant distances from the SMBH were likely launched from within the innermost few pc of the SMBH. In these cases, the gas is likely decelerating at those distances, implying that they had reached a peak velocity at closer distances. One input parameter to create the bicone models is the maximum projected velocity, $v\mathrm{_{max}}$. We limited the value of $v\mathrm{_{max}}$ to 600 km s$^{-1}$ so that there wouldn't be an inordinate amount of empty velocity model space that would come with generating models that had higher $v\mathrm{_{max}}$ values and which fit those discrepant points, but a bicone model that successfully fit those data could surely be generated.

Additionally, following the logic of rotational kinematics close to the minor axis given the orientation of the galactic disk, we expect to see small ($\sim50$ km s$^{-1}$) levels of redshifted rotational motion in the SW direction, as indicated by the green rotation curve plotted in Figure \ref{fig: STIS kinematics}. It is thus surprising to note the presence of blueshifted rotational motion in the SW regions of the G750M5, G750M6 and G750M7 plots. These kinematics are denoted by the blue data from $-$1\farcs2 $-$ 0$''$, which we attribute to rotation due to the relatively constant velocities across distance. One possible explanation for this discrepancy is that the circumnuclear molecular ring described in Section \ref{circumnuclear ring}, which is creating counterrotation in the inner 1$''$, may be affecting the motions of the ionized gas so that they are blueshifted.

\subsubsection{High-velocity Outflows and Asymmetries}
We see high-velocity outflows in the He~I and [S~III] emission, both of which are measured in the Z-band. In both cases, the outflows reach a velocity of $\sim$900 km s$^{-1}$ but stay fairly close to the SMBH, reaching distances of only $\sim0\farcs1$ (11 pc). Figures  \ref{fig: He I kinematics} and \ref{fig: SIII kinematics} show the spatial coverage of these outflows. \cite{barbosa09} also find evidence of high-velocity [S~III] outflows, but whereas we see them only in the redshifted direction, they find speeds of 900 km s$^{-1}$ in the blueshifted direction to the NW of the nucleus. They also find redshifted outflows in the area just south of the nucleus, reaching velocities of up to 500 km s$^{-1}$.
High-velocity outflows outside of the general flow pattern of ionized clouds are often seen close to the nucleus in the NLRs of nearby AGN, reaching a radial velocity of up to $\sim$2000 km s$^{-1}$ \citep{das05}, which provides an effective limit for our fits.

Another noticeable feature among the velocity maps for [S~III] (shown in Figure~\ref{fig: SIII kinematics}) is an asymmetry in the velocities, in the sense that the blueshifted and redshifted regions are not diametrically opposed with respect to the nucleus. We expect symmetry from gas whose motions are entirely rotational, so deviations are likely attributable to the outflows that are present in the inner arcsecond.
Asymmetrical outflows have been mentioned in previous studies of these NIFS observations \citep{riffel17, riffel21, schonell19, bianchin22}, as well as those based on other ground-based observations of NGC~3227 \citep{hicks08}. The [S~III] emission reflects the shape of the bicone in the NE quadrant (see Figure~\ref{fig:fig1}). The SW cone is revealed in these observations due to less extinction in the infrared by the host galaxy's disk, although extinction by dust lanes SW of the nucleus (see Figure \ref{fig: Judy Schmidt}) may still be responsible for its asymmetric location. The second, fainter kinematic component in the SW is slightly blueshifted, consistent with STIS data, although it's not clear why we do not detect a faint redshifted component in the NE.
\cite{barbosa09} note two clumpy blueshifted [S~III] regions, located 2\farcs4  and 0\farcs7 from the nucleus in the NE direction. They propose that these regions are caused by compression from a radio jet onto the interstellar medium (ISM), which is supported by the orientation of the radio contours from 3.6cm data from the Very Large Array \citep{nagar99, schmitt01}. Due to the smaller field of view, our NIFS data show only the clump at $\sim 0\farcs5$.

For Pa$\beta$, shown in Figure \ref{fig: PaB kinematics}, the outflows peak in velocity at 0\farcs3 to 0\farcs5 on either side of the nucleus, similar to that seen in the STIS data. \cite{schonell19} also finds a high concentration of blueshifted gas at this location in the NW direction, and \cite{bianchin22} see this gas in their map of the second kinematic component of Pa$\beta$. Overall, the velocity patterns, velocity dispersions, and fluxes of the ionized gas in the nuclear regions of NGC~3227 match those in other studies.

\begin{figure*}
\centering  
\subfigure[]{\includegraphics[width=0.45\linewidth]{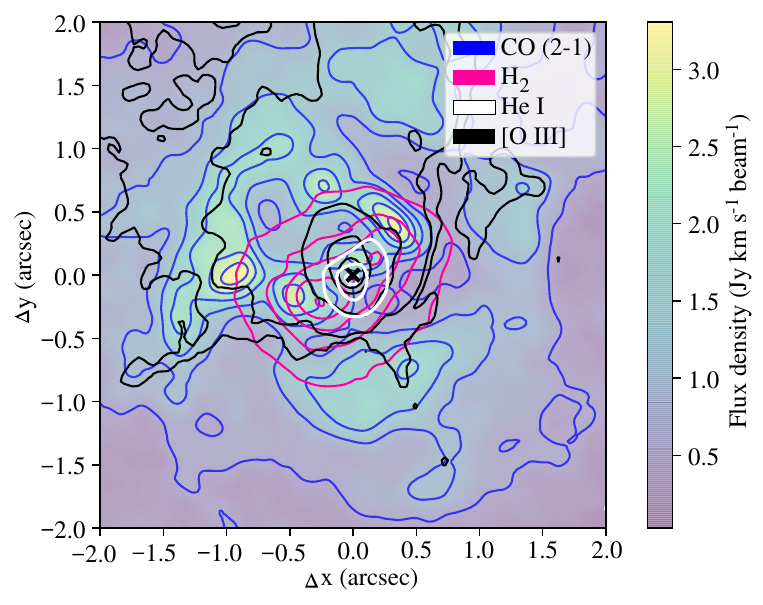}\label{fig: overplotted}}
\subfigure[]{\includegraphics[width=0.46\linewidth]{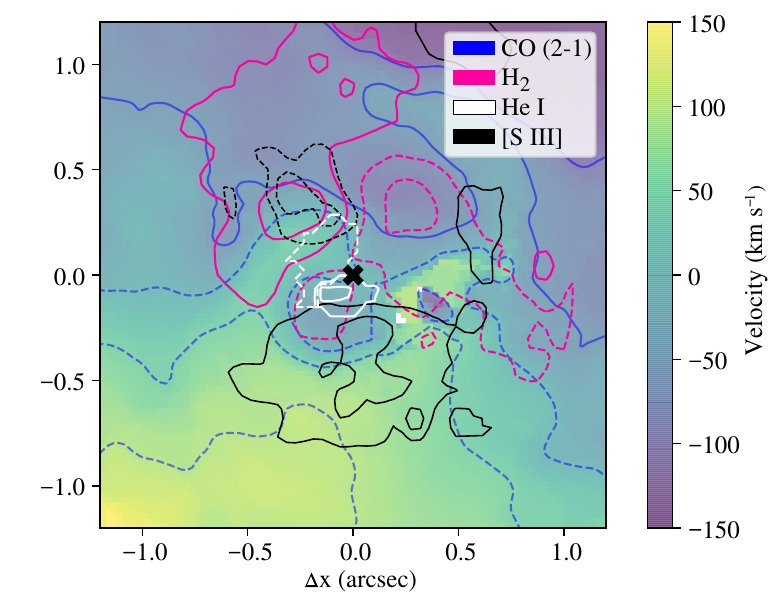}\label{fig: ALMA vels}}

\caption{Left: Flux contour plots using multiwavelength data including CO(2-1) \citep{alonso19}, [O~III] from the reduced HST WFC3 image, and He~I and H$_2$ from the NIFS Z- and K-bands, respectively. To create the flux contours for He~I, we decided to sort the three components in order of flux. Because the latter two components are significantly dimmer than the first, we opted to only show the contours for the first component. The background color map shows the flux density of CO(2-1). Right: Velocity plots showing the contour of He~I, H$_2$, and [S~III] outflows overlaid on to CO(2-1) velocity map from \cite{alonso19}. For the He~I velocity contours, we chose to sort the components by velocity, as is shown in Figure \ref{fig: He I kinematics}, and we plot all three components. The dashed contours show negative (blueshifted) velocities, while solid contours show positive (redshifted) velocities. It is important to note that although velocity due to stellar rotation has been subtracted from  He~I, H$_2$, and [S ~III], it has not been subtracted from the CO(2-1) emission. In both plots, the X at the center shows the location of the nucleus.}
\label{fig: flux contours}
\end{figure*}




%

\subsection{Multiphase Correlation}

We can characterize the kinematics to better understand how the various phases of gas that being studied -- ionized, warm molecular, and cold molecular -- interact with each other. \cite{fischer17} noted in their study of Mrk 573 that there is a progression of increasing distance and velocity from dust lanes, to H$_2$ emission, and finally to [S~III] emission, indicating in situ acceleration, heating, and ionization of the cold gas as the primary source of NLR outflows. A similar trend is found in Mrk~3 \citep{gnilka20}.

Figure~\ref{fig: flux contours} shows how the flux contours of CO(2-1), [O~III], H$_2$, and He~I significantly overlap with one another in the innermost couple of arcseconds in NGC~3227.
Going NE from the nucleus along the near cone, the above trend appears to hold as the extents and velocities of the gas increase with increasing ionization level. The only exception is the extent of the He~I recombination line, although its velocity follows the trend.

However, the opposite trend in level of ionization is seen along the bridge of gas about 1$''$ in extent, running SE to NW across the nucleus and capped by knots of CO(2-1) emission. We see a phase progression as we move in either direction along the bridge, starting with hot [O~III] and warm He~I emission in the center, moving outwards to cooler H$_2$ gas, which has its densest clumps $\pm$0\farcs25 (29 pc) from the center, and finally reaching the cool CO(2-1) clumps at $\pm$0\farcs5.  This progression indicates the hollowing out of the cold gas reservoir at the nucleus by ionizing radiation as suggested by \cite{alonso19}, which then flows along the cones.
Thus, the general trend of in situ heating, ionization, and acceleration of cold molecular gas to form the NLR outflows continues in NGC~3227, although at present most of the action appears to happen within $\sim$0\farcs5 (58 pc) of the nucleus.



We also study the velocities of the various phases to understand their interplay with one another. Figure~\ref{fig: H2 vel} shows how the velocities of CO(2-1), H$_2$, He~I, and [S~III] are correlated with one another. We notice several instances of overlap between positive and negative contours representing redshifted and blueshifted emission, respectively. We assume that based on the bicone structure observed in Figure~\ref{fig: bicone on WFC3 image}, in these instances, we are observing emission from both cones in the bicone structure. As a result, we will analyze the blueshifted and redshifted kinematics separately.


We see blueshifted outflows at 0\farcs4 NE in [S~III], which contain slight overlap with the edge of the He~I outflows. We see a progression in velocity as the He~I outflow reaches 300 km s$^{-1}$ on its northern end 0\farcs2 NE, after which it transitions to [S~III] emission that spans an additional 0\farcs4 NE at velocities of $\sim$100 km s$^{-1}$, until it finally leads into the surrounding CO(2-1) emission, which is moving at speeds of $\sim$70 km s$^{-1}$. One thing to note is that although we do see a velocity progression in the blueshifted emission, we do not see a corresponding phase progression.

The redshifted emission appears predominantly to the SW of the SMBH. The exception to this is the H$_2$ emission, which we are not heavily considering in the following commentary about outflows because we have established that its kinematics are primarily rotational (see Section  \ref{sec: H2 kin}). Much like the blueshifted outflows, the redshifted outflows follow the same spatial progression of transitioning from He~I to [S~III] to CO(2-1). The He~I outflows, located entirely within the hollowed CO(2-1) cavern, reach extraordinary speeds of 800 km s$^{-1}$ and extend SW towards the [S~III] outflows, which have  speeds of 100$-$150 km s$^{-1}$. It then disperses into the CO(2-1) emission, which has speeds of 75$-$125 km s$^{-1}$.

By comparing the kinematics of the CO(2-1), H$_2$, He~I, and [S~III] to one another, we see how the outflowing gas changes in phase and speed as the gas is accelerated away from the SMBH.  In both the redshifted and blueshifted directions, the He~I gas moves at high (250$-$800 km s$^{-1}$) speeds from the nucleus, until it interacts with slower (100$-$150 km s$^{-1}$) [S~III] gas 0\farcs2-0\farcs25 from the nucleus. Ultimately, the gas reaches the CO(2-1) emission, which is moving even slower (70$-$125 km s$^{-1}$). Thus, observing all these measurements together, we can observe how velocity tends to decrease with distance from the center.

\subsection{Influences from NGC 3226?}


Tidal influences on the AGN have been observed in other studies. \cite{fischer17} found that narrowband [O III] imaging of Mrk~573 shows the presence of a 3$''$ filament in the vicinity of the nucleus. They interpret the filament as a minor merger potentially undergoing tidal stripping as it infalls towards the AGN. Similarly, \cite{gnilka20} found that a large-scale gas/dust disk at the center of Mrk~3 is oriented towards a companion galaxy, which provides evidence for tidal fuelling of Mrk~3's AGN by that companion galaxy.

Seminal works from \cite{rubin68} and \cite{mundell95} have studied the tidal connections between NGC~3226 and NGC~3227 that have led to a richly intertwined dynamical history for the two galaxies. \cite{appleton14} note the complex structure of H~I and optical gas in this system comprising NGC~3226 and NGC~3227, also known as the Arp~94 system, and suggest that NGC~3226 may itself be the result of a recent merger, and may be sharing tidal debris from that event with NGC~3227. Nevertheless, in our KOSMOS measurements shown in Figure \ref{fig: KOSMOS kinematics}, which span almost the entire length of Arp~94, we see no sign of unusual kinematic signatures that would suggest a noteworthy tidal influence on the H$\alpha$ gas. We hypothesize that at distances close to the nucleus, any distinct kinematic signatures from the infalling gas are no longer distinguishable because integration into the disk of NGC 3227 has likely smoothed them out.


\section{Radiative Driving}
\label{radiativedriving}
In the NLR, photoionized gas is likely pushed away from the nucleus as a result of radiative driving mechanisms \citep{proga00, ramirez12, meena21}. In this case, two opposing forces control the velocity of the outflowing gas as a function of distance: AGN-induced radiative acceleration, and gravitational deceleration from the enclosed mass in the host galaxy, including the SMBH \citep{meena21}. The resulting equation, derived in full in \cite{das07}, is
\begin{equation}\label{eq: rad driving}
v(r)=\sqrt{\int_{r_1}^{r}\left[ 4885\frac{L_{44}\mathcal{M}}{r^2}-8.6 \times 10^{-3} \frac{M(r)}{r^2} \right]} dr
\end{equation}
where \textit{v(r)} is the outflow velocity (in km s$^{-1}$) of a cloud at a distance \textit{r} (in parsecs), L$_{44}$ is the bolometric luminosity in $10^{44}$ erg s$^{-1}$,  $\mathcal{M}$ is the force multiplier, and $r_1$ is the launch distance of the cloud from the central SMBH. For accelerated gas and dust, the force multiplier ($\mathcal{M}$) describes the ratio of the total absorption and scattering cross-section to the Thomson cross-section. $M(r)$ is in the units of solar masses (M$_\odot$). The first term in the equation is the accelerative force, while the second term is decelerative. Given the velocity and distance of a gas knot from the SMBH, which has been corrected for projection effects, the launch distance $r_1$ can be solved for numerically.
Given fixed values for the parameters in the above equation, we define the model turnover radius $r_t$ to be the distance from the SMBH at which the above two terms (radiative acceleration and gravitational deceleration) are equal, which is also equal to the maximum possible launch radius $r_1$ \citep{meena23}.

We can use the force multiplier $\mathcal{M}$ to specify how the efficiency of radiative driving from bound-bound, bound-free, and free-free mechanisms compares to that from pure Thomson scattering ($\mathcal{M}$ = 1) \citep{meena23, castor75, abbott82, crenshaw03, trindade21}. $\mathcal{M}$ is affected by factors such as the ionization parameter $U$ (which increases with decreasing $\mathcal{M}$ for typical NLR gas) and the spectral energy distribution (SED) \citep{trindade21}. We use a force multiplier $\mathcal{M}$ = 500, which was found to yield good agreement between modeled and observed turnover radii for a sample of 7 AGN in \cite{meena23}.


\subsection{Estimates of Bolometric Luminosity}
The AGN at the center of NGC~3227 has a variable luminosity, so the changing L$_{44}$ values will result in a range of possible turnover and launch radii. \cite{merritt22} finds a bolometric luminosity of $L_{\mathrm{Edd}} = 1.41 \times 10^{43}$ erg s$^{-1}$ from determination of the optical flux at 5100 \r{A}, which was corrected for reddening of the Seyfert 1 nucleus \citep[E$_{B-V}$ $=$ 0.18]{crenshaw01} and converted to bolometric luminosity through the calculation $L_{\mathrm{bol}} = 9.8 \lambda L_{5100}$ \citep{mclure04}. Assuming a constant SMBH mass of $M_{BH}$ = 1.1 $\times 10^7$ M$_\odot$ \citep{bentz23}, the maximum possible luminosity of the AGN in NGC~3227 is given by the Eddington limit as $L_{\mathrm{Edd}} = 1.4 \times 10^{45}$ erg s$^{-1}$. 

To obtain an average value of luminosity over the span of a decade, we retrieved fluxes at 5100 \AA\ from several observations of NGC 3227 \citep{denney10, derosa18, brotherton20}, including spectrophotometric observations we obtained from Lowell with 2$''$-wide slits on five occasions. All of these observations were obtained between 2007 and 2017, and the AGN flux at 5100 \AA\ varied by a factor of 2.2 during this time. After subtraction of the galaxy contribution from the above sources (6.8 $\times$ 10$^{15}$ erg s$^{-1}$ cm$^{-2}$), the same reddening and bolometric corrections as above, and using a distance of 23.7 Mpc \citep{tonry01, blakeslee01}, we determined an average bolometric luminosity of log $L_{\mathrm{bol}} = 4.78 \times 10^{43}$ erg s$^{-1}$ over 2007 -- 2017.

For our analysis we also wish to obtain an average value for the bolometric luminosity across the span of the NLR outflow, which has a light-travel time of at least 2000 years.
We therefore measured the integrated fluxes of [O~III] $\lambda$5007 in each of our 5 Lowell spectra obtained over 2009 -- 2010 to obtain an average value of 9.6 $\pm$ 2.3 $\times$ 10$^{-13}$ erg s$^{-1}$ cm$^{-2}$ \AA$^{-1}$.
Utilizing the relationship $L_{\mathrm{bol}}=3500L_{5007}$ \citep{heckman04}, we find an average bolometric luminosity of $L_{\mathrm{bol}} = 2.25 \times 10^{44}$ erg s$^{-1}$.  
Comparing the $L_{\mathrm{bol}}$ value calculated from [O~III] $\lambda$5007 emission, which represents an average luminosity over at least 2,000 years, to that calculated from 5100 \AA\ emission, which shows emission over the recent period of 2007 -- 2017, reveals that the brightness of the AGN has decreased by a factor of 8 over the last couple thousand years. Although the [O~III] $\lambda$5007 emission does not reveal specific temporal information to provide a timeline of when the dimming started, we can nevertheless confirm a substantial dimming of the NLR over this time.

\subsection{Launch Radii}
We calculate a mass profile using the same technique described in \cite{fischer17} and \cite{meena21}, which uses Equation~4 of \cite{terzic05} to calculate the mass density distribution for individual Sérsic components \citep{sersic68} that include the host galaxy disk and bulge. We use Equation~10 of \cite{terzic05} and the mass density distribution to determine the enclosed mass, and consequently gravitational deceleration, at a given radius from the SMBH. Based on the known values of the bolometric luminosity, enclosed mass and observed velocities at a observed distances (corrected for outflow projection), we can derive the outflow velocity profiles and numerically calculate the launch radii ($r_1$) of the observed outflowing gas using Equation~\ref{eq: rad driving}.

\begin{figure}[t]
  \centering
  \subfigure{\includegraphics[width=\columnwidth]{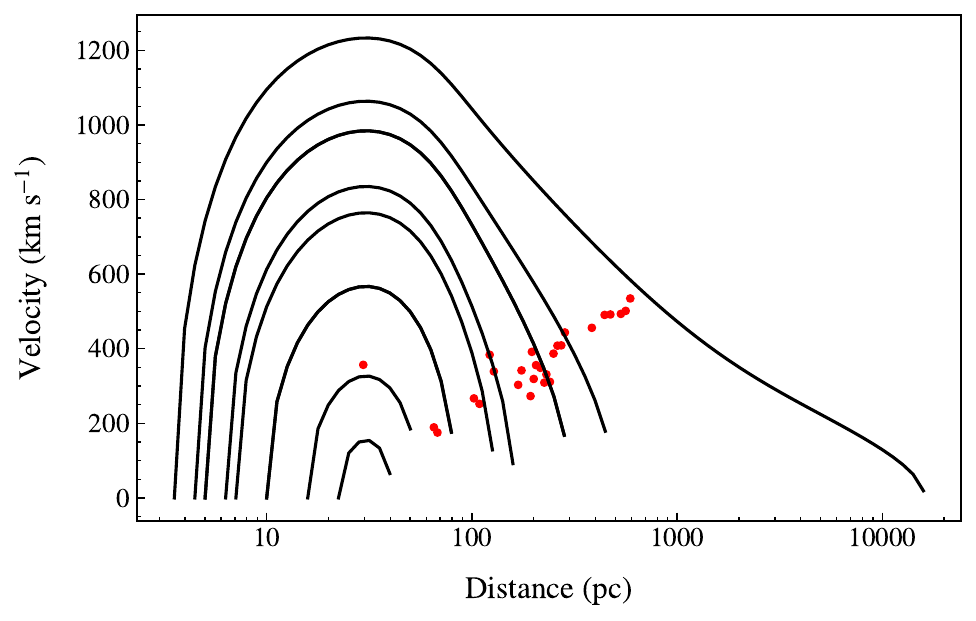}}
  \caption{Velocity profiles showing the trajectories of radiatively driven gas using Equation \ref{eq: rad driving}, a bolometric luminosity of L$_{bol}=2.25 \times 10^{44}$ erg s$^{-1}$ cm$^{-2}$, and a force multiplier of $\mathcal{M} = 500$. The red points represent kinematic data from KOSMOS, STIS, and NIFS whose distances and velocities have been deprojected using the bicone model. The turnover radius for these parameters is 31 pc.}
  \label{fig:vel profiles}
\end{figure}

Figure \ref{fig:vel profiles} shows how various launch radii will create different trajectories of radiatively driven gas, where gas launched from closer to the SMBH can be driven faster and to a greater distance than gas that is launched farther from the SMBH. Using the average bolometric luminosity over the NLR $L_{\mathrm{bol}} = 2.25 \times 10^{44}$ erg s$^{-1}$ cm$^2$, we find that the turnover radius for these models ranges from $31 - 63$ pc for force multipliers ranging from $\mathcal{M} = 500 - 3000$ (see more discussion of $\mathcal{M}$ values in \cite{meena23}). For comparison, the range of turnover radii is $13-28$ pc for the 2007 -- 2017 luminosity of 4.78 $\times 10^{43}$ erg s$^{-1}$ and $44 - 224$ pc for the Eddington luminosity. 

The data points in Figure \ref{fig:vel profiles} are distances and velocities of isolated emission-line knots in the KOSMOS, STIS, and NIFS datasets, corrected for projection effects with our bicone model.
These knots, at current distances of 30 -- 600 pc, can be tracked back to launch radii of 2 - 20 pc for the above parameters, with the upper end close to the maximum launch radius of 31 pc.

According to the best-fit kinematic model described in Section \ref{newmodel}, we find an observed turnover radius of $26^{+6}_{-6}$ pc from a purely empirical fit to the radial velocity profiles.
Based on our dynamical calculations of radiative driving by the AGN and gravitational deceleration from an enclosed mass model of NGC~3227, which are independent of the observed kinematics, we find a model turnover radius of 
$31 - 63$ pc for our NLR-averaged luminosity and force multiplier of $\mathcal{M} = 500 - 3000$.
Thus, we find remarkable agreement between observed and model turnover radii, indicating radiative driving and gravitational deceleration are the dominant dynamical forces in the NLR of NGC~3227. 

\begin{figure}[t]
\centering  
\includegraphics[width=\linewidth]{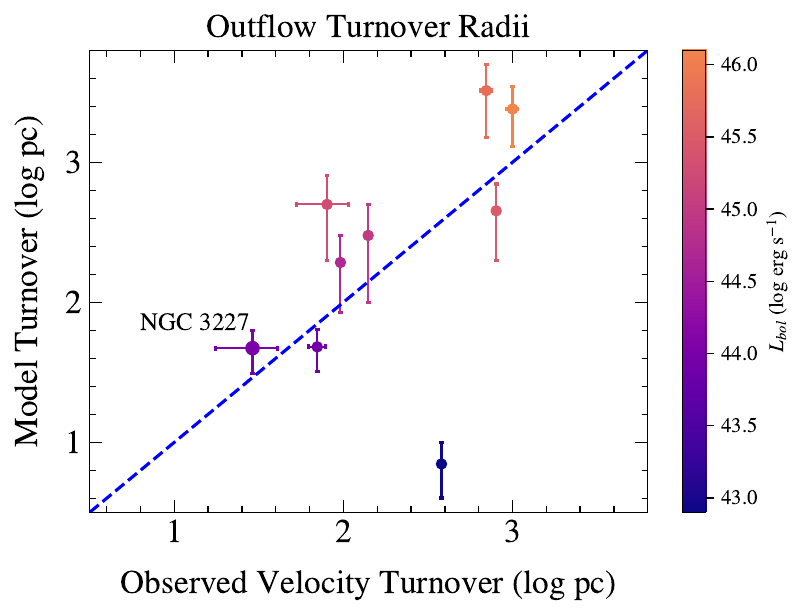}
\caption{Adapted from Figure 10 of \cite{meena23}. The model turnover radii obtained from radiative-gravity formulism (Section \ref{radiativedriving}) is compared to observed velocity turnover radii determined from the geometry of the biconal outflow model (Section \ref{newmodel}), but with the addition of NGC 3227. The colors of the points reflect the bolometric luminosity of each AGN taken from \cite{meena23}, underscoring their finding that brighter AGNs correspond to higher turnover radii. As in \cite{meena23}, vertical error bars for NGC 3227 arise from choosing lower and upper $\mathcal{M}$ values of 500 and 3000. Horizontal errors are associated with the resolution element of STIS, which was used for all measurements. The blue dashed line corresponds to unity, where the modeled and observed turnover radii are equal.} 
\label{fig: turnover comparison}
\end{figure}

\subsection{Comparison to Past Results}

\cite{meena23} examines the relationship between observed and model turnover radii for the NLRs in eight Seyfert galaxies and find a strong correlation, confirming that radiative driving alone is sufficient to produce the highly accelerated gas clouds. Figure \ref{fig: turnover comparison} shows the addition of NGC 3227 to Figure~10 of \cite{meena23}, providing another point at the low end of the luminosity range (spanning 10$^{43 - 46}$ erg s$^{-1}$). 
The point lies close to the dashed blue line showing unity between observed and modeled turnover radii, further extending the strong relationship between these two independently obtained values. As explained in \cite{meena23}, the one discrepant point is the low-luminosity AGN NGC~4051, and can be explained either by an additional driving mechanism or by the inability of current telescopes to resolve its turnover radius.
For NGC~3227, as for most of the other AGN in this plot, a force multiplier on the low end, at $\mathcal{M} = 500$, provides better agreement between model and observed turnover radii.

\section{Conclusions}

We observed the Seyfert 1 AGN NGC~3227 with the APO KOSMOS spectrograph and Gemini NIFS IFU in the Z-band, and used those data alongside archival data from HST, NIFS and ALMA to obtain ionized gas, molecular gas, and stellar kinematic maps of the circumnuclear region and host galaxy disk. Our main conclusions are as follows:
\begin{enumerate}
\item   We confirm the biconical outflowing structure of the ionized gas in the NLR of NGC 3227, and  improve upon the model established by previous works.. The bicone has an inclination of 40$^{+5}_{-4}$\arcdeg with respect to our line of sight, which is 25\arcdeg steeper than the value reported in \cite{travisthesis}. It has an inner half-opening angle of 47$^{+6}_{-2}$\arcdeg, which provides a Seyfert 1 view as expected, and an outer half-opening angle of 68$^{+1}_{-1}$\arcdeg . 
These values place the bicone axis close to the galactic plane, and explain the morphology of the NLR (especially the lack of [O~III] emission in the SW) as a result of extinction by dust in the plane. The dust can be seen as multiple lanes, arcs, and spirals on both nuclear and galactic scales, and is also responsible for reddening of the nucleus and NLR in the optical \citep{crenshaw01}.
\item We see ionized outflows traveling up to 500 km s$^{-1}$ at distances as large as 7$''$ (800 pc) from the SMBH. This is a significantly higher distance than was recorded in \cite{travisthesis}, and although the emission is weak, it extends beyond the modeled height of the bicone and is potentially disruptive to star formation.
Furthermore, we see disturbed ionized gas with large FWHM out to a distance of 15$''$ (1.7 kpc), which is not conducive to star formation. The effective radius of the bulge is approximately 300 pc \citep{bentz18}, and is unlikely to grow any further as long as it remains in this disturbed state.
\item We calculate the trajectories of radiatively driven gas from various launch radii. Using a NLR-averaged bolometric luminosity of $L_{\mathrm{bol}} = 2.25 \times 10^{44}$ erg s$^{-1}$ cm$^{-2}$, we find that outflows propelled by radiative driving have a turnover radius ranging from $31-63$ pc, depending on values for the force multiplier which we allow to range from $\mathcal{M} =$ 500 -- 3000. This range matches the value obtained from the bicone model, 26$^{+6}_{-6}$ pc, signaling agreement between our observed turnover radius and that from the radiative-driving formalism described in \cite{meena23}. In further agreement with \cite{meena23}, our findings support radiative driving plus gravity as the dominant force behind NLR outflows and additionally extend this result to the lower range of AGN luminosity exhibited by NGC 3227. In general, this formalism works best for a force multiplier of $\mathcal{M} =$ 500 .

\item We see a correlation among the several phases of gas that we have analyzed. Specifically, we note the presence of a bridge of gas centered around the nucleus and extending $\sim$0\farcs5 ($\sim$60 pc) in both NW and SE directions. This bridge is seen in He~I, H$_2$, and CO(2-1) and is perpendicular to the outflowing axis, which may be the original (and potentially still present) fueling flow to the AGN.  Along the direction of the bicone axis, we see [O~III], [S~III], He~I, H$_2$, and CO(2-1) emission with decreasing velocity amplitudes in both the redshifted and blueshifted directions, consistent with in situ heating, ionization and acceleration of gas from the cold gas reservoir.

\item Interestingly, our radiative driving calculations reveal that the launch distances of the existing NLR emission-line knots lie in the 2 -- 20 pc range. In fact, if we operate within model parameters constrained by mass and NLR-averaged luminosity, they cannot exceed 31 -- 63 pc. This places the launch radii of all of the NLR clouds within the bridge of gas seen in Figure \ref{fig: overplotted}, which we have argued is the source of both the outflows and likely the original fueling flow. 

\end{enumerate}

Further work will involve the use of CLOUDY \citep{ferland13} to make spatially-resolved mass loss profiles in the manner of \cite{revalski21, revalski22}. Together with spatially resolved maps of the cold gas, this will allow us to create mass evacuation timescales that will provide a better picture of how long it takes for the AGN in NGC 3227 to clear out surrounding material. Performing these calculations will allow us to make inferences about NGC 3227's ability to quench star formation in its central regions.

\acknowledgments
The authors would like to thank the anonymous referee for the constructive feedback. J.F. would like to thank Dr. Almudena Alonso-Herrero for graciously sharing her ALMA data on NGC 3227. J.F. would also like to thank Dr. Jan Tobochnik for his helpful discussions regarding how to calculate the best bicone model. R.A.R. acknowledges the support from Conselho Nacional de Desenvolvimento Cient\'ifico e Tecnol\'ogico (CNPq; Proj. 303450/2022-3, 403398/2023-1, \& 441722/2023-7), Funda\c c\~ao de Amparo \`a pesquisa do Estado do Rio Grande do Sul (FAPERGS; Proj. 21/2551-0002018-0), and CAPES (Proj. 88887.894973/2023-00).

Many of the data presented in this work are based on observations with the NASA/ESA Hubble Space Telescope and were obtained from the Mikulski Archive for Space Telescopes (MAST), which is operated by the Association of Universities for Research in Astronomy, Incorporated, under NASA contract NAS5-26555. These observations are associated with program number \href{https://archive.stsci.edu/proposal_search.php?mission=hst&id=16246}{16246}. Support for program number 16246 was provided through a grant from the STScI under NASA contract NAS5-26555. The specific observations used in this work analyzed can be accessed via DOI:  \dataset[10.17909/cm86-me24]{\doi{10.17909/cm86-me24}}.

This research has made use of NASA’s Astrophysics Data System. IRAF is distributed by the National Optical Astronomy Observatories, which are operated by the Association of Universities for Research in Astronomy, Inc., under cooperative agreement with the National Science Foundation.
This research has made use of the NASA/IPAC Extragalactic Database (NED), which is operated by the Jet Propulsion Laboratory, California Institute of Technology, under contract with the National Aeronautics and Space Administration.
Some of the observations used in this paper were obtained with the Apache Point Observatory 3.5-meter telescope, which is owned and operated by the Astrophysical Research Consortium.
\clearpage







\bibliographystyle{aasjournal}
\bibliography{bibbo.bib} 

\restartappendixnumbering


\end{document}